\documentclass[a4paper,fleqn]{cas-sc}

\usepackage[authoryear]{natbib}

\usepackage{amsmath,amssymb}

\usepackage{subcaption}
\usepackage{booktabs}
\usepackage{array}
\usepackage{siunitx}
\usepackage{xcolor}
\usepackage{placeins}

\usepackage{listings}
\usepackage[capitalize,nameinlink,noabbrev]{cleveref}

\ifdefined\reviewmode
  \usepackage{lineno}
  \linenumbers
\fi

\newcommand{\executecode}{\texttt{execute\_code}}
\newcommand{\executetask}{\texttt{execute\_task}}
\newcommand{\interrupttask}{\texttt{interrupt\_task}}
\newcommand{\checktaskstatus}{\texttt{check\_task\_status}}
\newcommand{\listtasks}{\texttt{list\_tasks}}

\graphicspath{{figures/}}

\makeatletter
\AtEndDocument{\clearpage\immediate\write\@auxout{\string\csxdef{lastpage}{\the\numexpr\value{page}-1\relax}}}
\makeatother

\begin{document}
\let\WriteBookmarks\relax

\shorttitle{Connecting AI agents to live DEM simulations}
\shortauthors{Y.~Han et~al.}

\title[mode=title]{Connecting AI Agents to Live Discrete Element Simulations via the Model Context Protocol}

\author[1]{Yusong Han}[orcid=0000-0002-5453-2470]
\cormark[1]
\ead{han.yusong.53f@st.kyoto-u.ac.jp}
\credit{Conceptualization, Methodology, Software, Investigation, Visualization, Writing -- Original draft}

\author[2]{Ryosuke Uzuoka}[orcid=0000-0002-1543-1626]
\ead{uzuoka.ryosuke.6z@kyoto-u.ac.jp}
\credit{Supervision, Writing -- Review \& Editing, Funding acquisition}

\author[2]{Kyohei Ueda}[orcid=0000-0001-6202-6431]
\ead{ueda.kyohei.2v@kyoto-u.ac.jp}
\credit{Supervision, Writing -- Review \& Editing, Funding acquisition}

\affiliation[1]{organization={Graduate School of Engineering, Kyoto University},
            addressline={Gokasho},
            city={Uji, Kyoto},
            postcode={611-0011},
            country={Japan}}

\affiliation[2]{organization={Disaster Prevention Research Institute, Kyoto University},
            addressline={Gokasho},
            city={Uji, Kyoto},
            postcode={611-0011},
            country={Japan}}

\cortext[cor1]{Corresponding author}

\begin{abstract}
This paper describes a Model Context Protocol (MCP) server through which a general-purpose AI agent operates a discrete element engine, commercial or open-source, while the simulation runs. The agent submits a long simulation to the engine and, while it computes, reads the running model, changes it, or stops it. The human modeler stays at the engine's console and graphical interface throughout, and the agent sees what the modeler does there. The tools are derived from what the agent needs to do at the engine and are held to four principles: correctness, coverage, orthogonality and simplicity. Two documentation tools let the agent search the engine's manual and browse it level by level. Five execution tools let the agent run code in the engine, submit long simulations in the background, check on them, read the running model, and stop them. A bridge inside the engine serves these tools in a commercial DEM engine, PFC, and an open-source one, YADE. In a case study, a general-purpose agent calibrated a drained triaxial model of dense Toyoura sand against laboratory data in both engines, while the modeler worked at the same engine; the agent noticed the modeler's actions at the console and acted on them. The same calibration on YADE was then given to seven language models, three sessions each and without human intervention. Twenty of the twenty-one sessions reached the targets, and nineteen of the parameter sets lie in one small region of the admissible ranges. The server and the bridge are released as two open-source packages, itasca-mcp, which also serves FLAC, 3DEC, MPoint and MassFlow, and yade-mcp.
\end{abstract}

\begin{keywords}
Discrete element method \sep AI agents \sep Model Context Protocol \sep
Human--computer interaction \sep Calibration \sep Granular materials
\end{keywords}

\maketitle

\section{Introduction}
\label{sec:introduction}

The discrete element method (DEM) \citep{cundall1979} represents soil as an assembly of interacting particles and gives numerical access to the micromechanics behind its macroscopic behavior. Its use in practice, however, involves recurring difficulties. The microscopic parameters of a DEM model describe the particles and their contacts, and are different in kind from the macroscopic properties measured in the laboratory. No direct mapping leads from one to the other, so the parameters are found by calibration. Over repeated trials, the simulated response is compared with the measured laboratory data, and the parameters are adjusted until the simulation reproduces the laboratory results \citep{coetzee2017,belheine2009}. DEM is also computationally intensive: the forces at tens of thousands of contacts are computed at every step, and a single simulation can last hours or days. Throughout, a human modeler monitors the simulation, interrupts it when something goes wrong, corrects the script, and starts it again, until the task is complete. Each of these steps calls for judgment and is repeated many times. This raises a natural question: could an artificial intelligence take on part of this work, or even simulate the way the modeler does?

\textbf{Data-driven artificial intelligence.} Most applications of artificial intelligence in geotechnical engineering are data-driven: a model is trained on data and then predicts, and the agenda of data-centric geotechnics would extend this to an artificial intelligence that learns from the experience of the whole profession \citep{phoonzhang2023future}. The data come from every scale of the discipline. At the scale of the site, a model trained on site-investigation data predicts the soil profile \citep{sheil2026,phoon2022sitechar}. At the scale of the specimen, data-driven models of soil behavior learn the stress--strain response from element tests \citep{zhang2021mlconstitutive}. At the scale of the grain, the response of sand depends on the fabric of the assembly \citep{dafalias2004fabric,li2012acst}. Laboratory tests rarely measure the fabric, but DEM gives it directly and has linked it to macroscopic behavior \citep{zhao2013fabric}.

DEM simulations have become a source of training data. Some studies reviewed by \citet{zhang2021mlconstitutive} take the fabric from them as training data. Others calibrate the DEM model with a surrogate \citep{benvenuti2016,rackl2017}. They run many DEM simulations with different parameters and train the surrogate on the results, so that the surrogate predicts the macroscopic quantities from the parameters far faster than a DEM simulation. The DEM parameters that reproduce the experiment are then found by trying them on the surrogate. A third approach runs the DEM simulations themselves in rounds, with a Bayesian filter narrowing the parameters after each round \citep{cheng2019bayesian}. The surrogate and the Bayesian filter both need many DEM simulations, and the parameters they return must still be checked in further DEM simulations. Both steps still depend on a person to set up, check and judge the simulations. Data-driven artificial intelligence has thus taken on part of the work, the prediction; the work at the engine remains with the human modeler.

\textbf{From conversation to agent.} Large language models changed the question from what a model can predict to what a model can do. A language model generates text, including program code, and geotechnical researchers began to test it on their own problems. \citet{chatgptmatlab2024} document an early test in instructive detail: the engineer describes an engineering problem in a conversation with a chatbot and receives MATLAB code in return. The engineer then runs the code by hand, and comes back to the conversation to repair what failed. The study shows that a language model can write working numerical code for geotechnical problems, but the steps outside the conversation remain with the engineer. In the same period, \citet{parrottales2024} took the step beyond the conversation. A language model was given a set of geotechnical calculation tools, a bearing-capacity function among them, and solved a foundation problem by choosing which tool to call and reading what each returned. This combination of a language model with tools is what is now called an AI agent. An AI agent, in the sense used throughout this paper, is a program that runs a language model in a loop. The language model is given a goal and a set of tools it may call. The language model chooses a call, the program carries it out and returns the result, and the language model chooses again until the goal is met.

\textbf{Agent frameworks.} Geotechnical agents are now being built. GeoLLM and its multimodal successor orchestrate geotechnical design tasks through retrieval-augmented prompting, self-review agents, and human-in-the-loop checkpoints \citep{geollm2025,multigeollm2025}; agent-orchestrated frameworks give natural-language access to large site-investigation databases \citep{pan2026sitechar}; GeoSim.AI carries the assistant to the simulation software itself, wrapping the script syntax of commercial codes in retrieval and generating input files on demand \citep{geosimai2025}. International hackathons on agentic AI have followed. The first GeoTechathon set its teams on geological modeling, slope stability, liquefaction and tunneling \citep{wu2025perspectives}. At the second, one team calibrated constitutive-model parameters from laboratory curves, and another had an agent prepare the input of slope-stability software from site data \citep{wu2026geotechathon}. Outside geotechnics, agents operate the simulation software of neighboring fields. In solid mechanics, a team of agents writes, runs and corrects finite element code for elasticity problems \citep{mechagents2024}; in fluid dynamics, agents set up and run OpenFOAM cases \citep{openfoamgpt2025,foamagent2025}; and in structural engineering an agent operates the structural analysis software ETABS, from model setup to post-processing \citep{chen2026structural}.
These systems advance the agent itself, through retrieval over documentation, multi-agent orchestration and domain knowledge; none of them connects the agent to a running simulation.

\textbf{Challenges in connecting AI agents to a live engine.} The work reviewed above advances the agent and leaves the simulation engine as it was: the agent starts a simulation and waits for the simulation to end. During the simulation, the agent cannot see how the simulation is going, or stop the simulation, change the model and continue. This paper takes the opposite direction: it treats the agent as a user with needs and asks what tools the engine should offer it. Designing for this user meets four challenges:
\begin{itemize}
  \item \emph{Interfaces designed for human users}: the engine was designed for a modeler at a graphical interface and a command line, so an agent can neither type at the console or click in the graphical interface as the modeler does, nor see what either shows.
  \item \emph{Hallucination}: an agent may misremember the syntax of the engine, because the engine's commands appear only rarely in the training data of language models.
  \item \emph{Long-running simulations}: a DEM simulation can last hours. During the simulation, the agent has to follow the simulation and act on the model, and the engine offers the agent no interface for this.
  \item \emph{The human in the loop}: the human modeler and the agent share one engine, so each needs to see what the other does. The modeler must not only read the agent's reports but also check the model directly in the engine, watching the model in its graphical interface and verifying numbers at its console. Conversely, the modeler can also make mistakes at the engine and damage the model, so the agent must know what the modeler has done there. Human--computer interaction in both directions is what keeps the human in the loop.
\end{itemize}
General-purpose agents already write and debug code well; what they lack is not intelligence but access. Tools that meet these challenges, and an engine that serves them, form the infrastructure beneath the orchestration layer of AI agent frameworks.

\textbf{Building the interface on the Model Context Protocol.} A standard for such tools exists. The Model Context Protocol (MCP) \citep{mcp2024} standardized how agents discover and call external tools: a tool built to this standard is designed once, with the agent as its user, and any agent that supports the protocol can call it. The agent may come from any vendor, or be a language model run on the laboratory's own computers where project data are confidential.

Four MCP servers for engineering computation have been reported. One, GeoMCP, serves analytical geotechnical methods: each method is stored as a declarative card with its equations, units and applicability limits, and a symbolic engine evaluates the card \citep{geomcp2026}. Another wraps the command-line utilities of the traffic simulator SUMO and runs them one batch at a time \citep{sumomcp2025}. A third exposes the multiphysics package COMSOL through 76 tools covering the modeling workflow, with solves run in the background and documentation retrieved by semantic similarity from a vector database \citep{comsolmcp2026}. The fourth drives the geotechnical finite-element package PLAXIS 3D through 19 typed tools over the package's own scripting interface, with an orchestrator that runs a geometry, a calculation and an independent verification agent in turn, and builds staged tunnel models with no human operating the interface \citep{duangsano2026}. In all four, the agent submits the computation and waits for the result. Beyond this, the agent can at most look up methods or documentation, track the progress of the computation, or cancel it. While the computation runs, the agent can neither read nor act on the model, and the agent and the modeler do not see what the other does in the engine. \Cref{tab:landscape} summarizes the approaches reviewed above and where this work stands among them.

This paper builds such an interface for DEM, so that the agent works on the same running engine process as the modeler rather than alone. The tools are derived from the needs of the agent, which follow the working loop of the modeler (\cref{sec:interface,fig:architecture}). Two documentation tools let the agent search and browse the engine's manual, so that the code it writes follows what the engine documents (\cref{sec:docs}). The hard part is not reaching the engine but reaching it during a simulation: a long simulation must not keep the agent from reading the model, a read must not disturb the simulation, and a stopped simulation must leave the model intact (\cref{sec:bridge,sec:semantics}). To our knowledge, no interface has previously given an agent this access to a live DEM simulation. Such access also serves the data-driven work reviewed above: a surrogate needs many DEM simulations, and an agent that operates the engine can prepare, run and check the simulations \citep{benvenuti2016}.

\begin{table}[pos=t]
\caption{Artificial intelligence and the simulation engine in geotechnical engineering: the approaches and their workflows.}
\label{tab:landscape}
\small
\begin{tabular}{@{}>{\raggedright\arraybackslash}p{0.19\linewidth}>{\raggedright\arraybackslash}p{0.24\linewidth}>{\raggedright\arraybackslash}p{0.51\linewidth}@{}}
\toprule
Approach & Examples & Workflow \\
\midrule
\multicolumn{3}{@{}l}{\emph{Data-driven models}} \\
\addlinespace
Trained on laboratory or field data & \citealp{phoon2022sitechar,zhang2021mlconstitutive} & A model trained on measured data predicts soil properties or soil response. \\
\addlinespace
Trained on DEM simulations & \citealp{benvenuti2016,rackl2017,cheng2019bayesian} & A model trained on DEM simulations maps microscopic parameters to the macroscopic response and returns calibrated parameters; the modeler checks them in further simulations. \\
\midrule
\multicolumn{3}{@{}l}{\emph{Language models}} \\
\addlinespace
Chatbots for numerical code & \citealp{chatgptmatlab2024} & The chatbot writes numerical code on request, such as MATLAB scripts for seepage and slope stability; the engineer runs it by hand and carries the errors back to the conversation. \\
\addlinespace
Agents for design and data & \citealp{geollm2025,multigeollm2025,pan2026sitechar} & The agent extracts design information from documents or queries site databases, and carries out the calculations. \\
\addlinespace
Agents for simulation software & \citealp{geosimai2025,mechagents2024,openfoamgpt2025,foamagent2025,chen2026structural} & The agent generates the input, script or software commands; most agents also run them and correct them from the output. \\
\addlinespace
MCP servers for engineering computation & \citealp{geomcp2026,sumomcp2025,comsolmcp2026,duangsano2026} & The agent submits the computation through tools and waits for the result. Beyond this, the agent can at most look up methods or documentation, track the progress of the computation, or cancel it. \\
\addlinespace
This paper & & The agent submits a task, then reads and acts on the model while the task runs (\cref{sec:interface}); the agent and the modeler collaborate in one shared, live engine. \\
\bottomrule
\end{tabular}
\end{table}

\textbf{Outline.} The rest of the paper is arranged as follows. \Cref{sec:server} designs the tools and the bridge that serves them inside two DEM engines, PFC \citep{itasca2021}, which is commercial and closed, and YADE \citep{kozicki2008,yade2021,yade2024}, which is open-source; the two share one tool design, and only the part inside the engine differs. The two realizations are released as the open-source packages itasca-mcp and yade-mcp. \Cref{sec:casestudy} gives a general-purpose agent the calibration of a drained triaxial model of Toyoura sand in both engines, with the modeler at the engine throughout. \Cref{sec:limitations} states what the interface requires of an engine and what the calibration task did and did not test, and \cref{sec:conclusions} concludes.
 \section{The MCP server: from agent needs to a live engine}
\label{sec:server}

\begin{figure}[pos=t]
  \centering
  \includegraphics{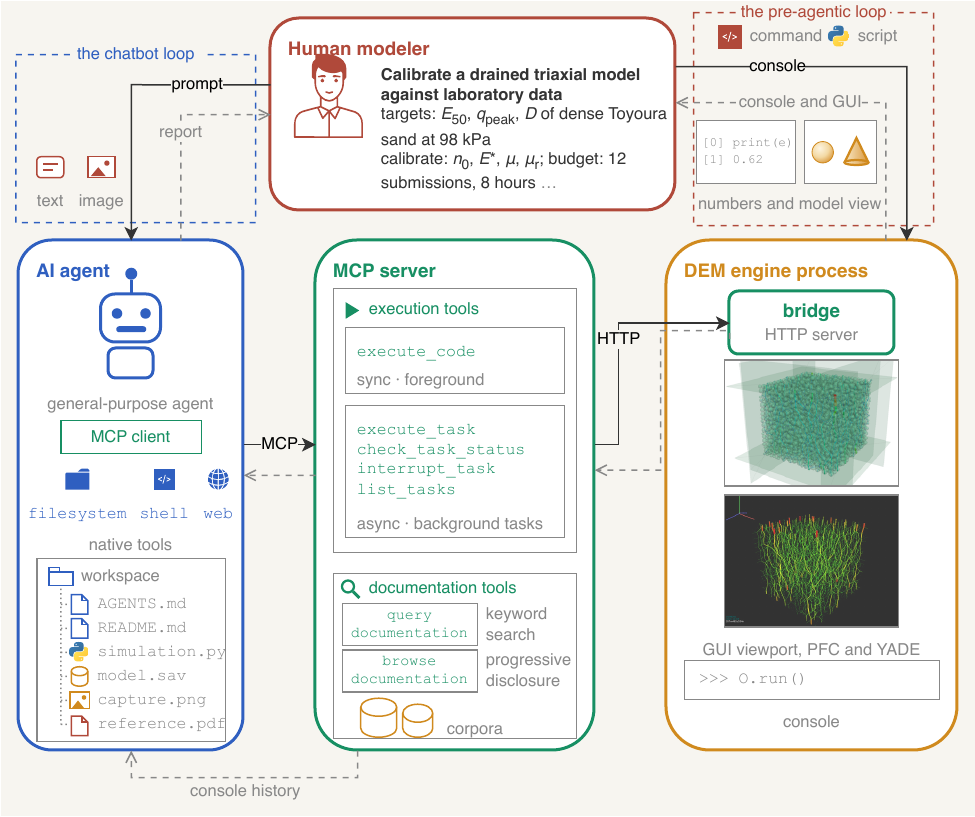}
  \caption{Architecture of the agentic loop. The modeler operates and reads the engine through its console and GUI; the AI agent operates and reads the same engine process through the MCP server. The modeler and the agent communicate in natural language. The modeler sees what the agent does to the engine in the console and the GUI, and the agent sees what the modeler does to the engine in the console history. The dashed frames mark the pre-agentic loop (top right) and the chatbot loop (top left). The MCP server and the bridge, drawn in green, are the contribution of this paper; the agent, its language model and the DEM engine are existing and are used as they are.}
  \label{fig:architecture}
\end{figure}

\textbf{The pre-agentic loop.} Consider how a DEM model is built without an AI agent. The modeler opens the manual to find the right command or API call, and checks its syntax. The command goes into the DEM engine's console. The modeler looks at the graphical user interface (GUI) to see what the command did to the particles and walls, and reads the numbers printed in the console to see what the command did to the state. The modeler then judges whether the result is as intended. If something looks wrong, the modeler adjusts the command and sends it again. Once every command does what it should, the modeler collects them into a script and submits the script to the engine for the full simulation, which may take hours or days. Look up, try, look, adjust, and launch the full simulation only once every command works. This loop is the daily practice of every DEM modeler. It is the pre-agentic loop in the top right corner of \cref{fig:architecture}, framed in dashed red: a command or a script goes in, the numbers and the model view come back.

\textbf{The chatbot loop.} A modeler with a question about the simulation can put the question to a chatbot in a conversation. The chatbot answers from the training data of its language model and returns a report or a script \citep{chatgptmatlab2024}. The modeler carries the commands or the script the chatbot wrote to the engine by hand, runs them, and comes back with the next question. This is the chatbot loop in the top left corner of \cref{fig:architecture}, framed in dashed blue. The chatbot never touches the engine and knows the model only from the conversation, so its answers are unreliable. The modeler still looks, reads, judges, and adjusts exactly as in the pre-agentic loop, and in addition carries text from the engine to the chatbot by hand.

\textbf{The AI agent.} Without the modeler relaying, the chatbot loop cannot run on its own. Automating the chatbot loop means letting the language model send the commands to the engine and read the results itself, and the mechanism for that is the tool. A tool is an action the language model is allowed to request. The declaration of a tool contains a name, a description of the action, the parameters of the action, and a description of each parameter. The tools are given to the language model together with the modeler's request. The language model decides which tool to call and with which arguments. The system performs the action and returns the result, and the language model reads the result and decides what to do next. A general-purpose agent, the bottom left box of \cref{fig:architecture}, is a program that runs a language model with tools in this way. The agent works in a folder on the modeler's computer, the workspace, which holds the scripts, saved models, captured images, and reference papers. The native tools of the agent read and write files and run shell commands.

\textbf{The agentic loop.} Those native tools are enough to write a simulation script, but not to run it: none of them reaches into the engine's process. What is needed is a set of tools built for the engine. Every language model provider describes and calls tools in its own format, so a tool written for the language models of one provider has to be rewritten for the next. The Model Context Protocol removes that problem: it is a shared standard by which an agent connects to a server, learns what tools it offers, and calls them, so a tool is written once and any agent that supports the protocol can use it. This paper contributes such a server and a bridge for it, the MCP server and the bridge drawn in green in \cref{fig:architecture}. The MCP server, in the middle of the figure, offers the tools to the agent. The bridge, in the box of the engine, receives the calls from the MCP server inside the engine's process. The agent, its language model and the DEM engine are existing software, and this paper does not change them. Through the MCP server and the bridge the agent works on the same running engine process as the modeler, and reaches the engine through the tools the server offers. The agent now looks, runs, reads, and adjusts on its own, as the modeler does in the pre-agentic loop. The modeler keeps the chatbot loop, asking and judging, and keeps the pre-agentic loop as well, working at the console and watching the GUI. The modeler stops relaying. Human--computer interaction runs in both directions: the modeler sees what the agent does to the engine in the console and the GUI, and the commands the modeler types at the console return to the agent as console history. That is the agentic loop: the modeler and the AI agent working on one engine process. \Cref{tab:loops} lists who carries out each step in the three loops. \Cref{sec:interface} explains why the MCP server offers these tools, and \cref{sec:docs,sec:bridge,sec:semantics} explain why the tools are designed in this way.

\begin{table}[pos=htbp]
\caption{Who carries out each step of the work at the engine in the three loops of \cref{fig:architecture}.}
\label{tab:loops}  \begin{tabular}{@{}>{\raggedright\arraybackslash}p{0.23\linewidth}>{\raggedright\arraybackslash}p{0.19\linewidth}>{\raggedright\arraybackslash}p{0.25\linewidth}>{\raggedright\arraybackslash}p{0.24\linewidth}@{}}
\toprule
Step & Pre-agentic loop & Chatbot loop & Agentic loop \\
\midrule
Set the task & Modeler & Modeler & Modeler \\
Look up the command & Modeler, in the manual & Chatbot, from the training data of its language model & Agent, through the documentation tools \\
Try the command, adjust it, and run the full simulation & Modeler, at the console & Modeler, who carries text between the chatbot and the console by hand & Agent, through the execution tools; the modeler can also, at the console \\
Read the result & Modeler & Modeler & Agent and modeler \\
Accept the result & Modeler & Modeler & Modeler \\
\bottomrule
\end{tabular}
\end{table}

\subsection{From agent needs to tools}
\label{sec:interface}

\textbf{Four principles.} The opening of this section described how the agent reaches the modeler's engine process through the tools the server offers. This subsection designs those tools, starting from what the agent needs to do. Four principles guided the design of the tools:
\begin{itemize}
  \item \emph{Correctness}: the interface must tell the truth about the simulation.
  \item \emph{Coverage}: the interface must serve each thing the agent needs to do.
  \item \emph{Orthogonality}: the interface must offer operations that overlap as little as possible.
  \item \emph{Simplicity}: the interface must meet the first three with as little machinery as possible.
\end{itemize}
The agent's needs set what the tools are to do, and the four principles constrain how the tools are built.

\textbf{The first need: a check on the commands.} PFC's command language runs to thousands of commands, YADE's Python API documents several hundred classes, and both change between versions, so the modeler does not memorize the commands and checks each one in the manual. A language model cannot write the commands from its training data either: a command written from the training data may not exist in the engine, or may mean something else. The agent needs the same check as the modeler, and that is the first need. The first principle, \emph{correctness}, requires it.

\begin{figure}[pos=t]
  \centering
  \includegraphics{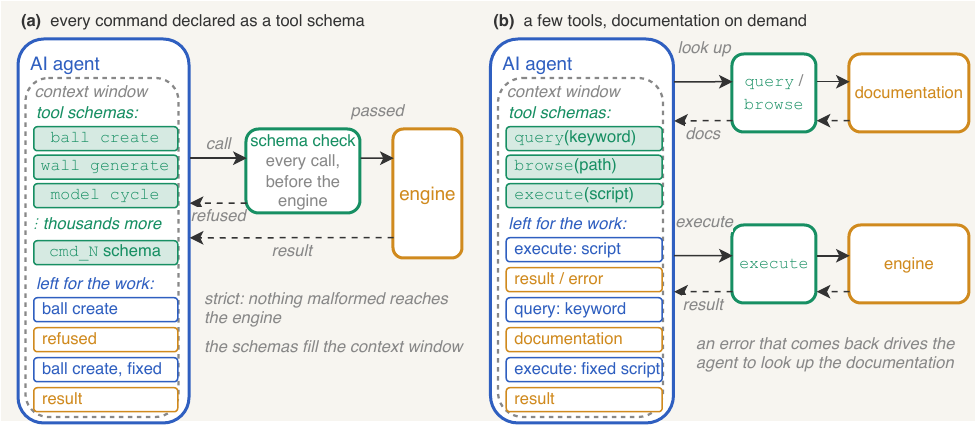}
  \caption{Two ways to check the commands an agent writes. (a) Every engine command declared as a tool schema: each call is checked against its schema before it reaches the engine, and every schema occupies part of the language model's context window. (b) The route taken in this study: the agent looks the command up with two documentation tools before writing it. A skipped lookup is caught by the engine's own error message, which drives the agent to look the command up.}
  \label{fig:docloop}
\end{figure}

MCP offers one way to check: every call is checked against the schema of the tool, the part of the declaration that lists the parameters, so declaring every engine command as a tool would check every command. But the input of a language model has an upper limit, the context window, which is measured in tokens, and the declarations of all the tools go into the window in full. Thousands of declarations would fill the window (\cref{fig:docloop}(a)). The server therefore lets the agent look the command up, as the modeler does. Two documentation tools replace the thousands of declarations: a query tool searches the documentation by keyword, and a browse tool returns the documentation at a given path. The browse tool reaches every command in the documentation, which meets the second principle, \emph{coverage}. Only the schemas of these two tools occupy the context window (\cref{fig:docloop}(b)). \Cref{sec:docs} describes the two tools.

\textbf{The second need: a console.} With the command looked up, the modeler tries it in the console. The engine accepts the command or rejects it. The modeler reads the effect of the command in the GUI and in the console (e.g., the velocity of a particle or the stress on a wall). The agent needs the same console: a tool that takes a code snippet, runs it in the engine, and returns the engine's response unaltered. This is \executecode{}, the first tool in the green box of \cref{fig:architecture}, the console as a tool. When the engine rejects a command, \executecode{} returns the error message of the engine, and the error drives the agent to look the command up in the documentation (\cref{fig:docloop}(b)).

Once each command has been looked up, tried, and checked in the console and the GUI, the modeler collects the commands into a script and submits the script to the engine for the full simulation. The simulation takes hours or days, and the modeler needs to check the simulation while it runs. The modeler reads the output the script prints in the console, and also watches the model in the GUI to see whether the simulation runs as intended. When the output or the model view looks wrong, the modeler interrupts the simulation, fixes the script and submits it again, until the simulation finishes.

The agent has tried the code through \executecode{} and now has a script to run as well. The agent writes the script with the filesystem tool into the workspace shown in the AI agent box of \cref{fig:architecture}. Like the modeler, the agent needs to read the output of the simulation while it runs. When something looks wrong, the agent interrupts the simulation and fixes the script. The agent then resubmits the script, and repeats these steps until the simulation finishes. The agent therefore needs four background tools. \executetask{} submits the script as a task and returns at once with the identifier of the task. \checktaskstatus{} takes the identifier of a task, reports whether the task is running, finished or failed, and returns what the task has printed. \interrupttask{} stops the task without closing the engine process, so the DEM model stays in the engine at the step where the task stopped. \listtasks{} returns the task history of the workspace, every past task and how it ended. \checktaskstatus{} accepts the identifier of a past task as well, so the work done in the engine can be traced back task by task. The output of the script is not always enough. For example, to see whether the simulation is numerically stable, the agent reads the largest contact force in the model and captures the model view in the GUI as an image. A script usually reports neither, and a script cannot report everything the agent may want to check. For this, \executecode{} remains available while a task runs, and the agent uses it to read any quantity or capture any view of the running simulation. How the engine answers such a read while it cycles is described in \cref{sec:bridge,sec:semantics}.

\begin{table}[pos=htbp]
\caption{The tools the MCP server offers the agent, as drawn in the MCP server box of \cref{fig:architecture}.}
\label{tab:tools}
\begin{tabular}{@{}lp{0.72\linewidth}@{}}
\toprule
Tool & Description \\
\midrule
\texttt{query documentation} & Searches the documentation by keyword and returns the paths of the matching commands. \\
\texttt{browse documentation} & Returns the full documentation of a command, or the list of commands below a path. \\
\addlinespace
\executecode{} & Runs a code snippet in the engine and returns the engine's response. Remains available while a task runs. \\
\executetask{} & Submits a script as a background task and returns the task identifier. \\
\checktaskstatus{} & Reports the state of a task and returns its output so far. \\
\interrupttask{} & Stops a running task and keeps the state of the engine. \\
\listtasks{} & Lists the task history of the workspace, every past task with its final state. \\
\bottomrule
\end{tabular}
\end{table}

\textbf{Summary.} \Cref{tab:tools} lists the tools: two documentation tools and five execution tools. The commands the agent writes are checked against the engine's own documentation, which meets \emph{correctness}. The browse tool reaches every command and \executecode{} runs every command the console accepts, which meets \emph{coverage}. \executecode{} serves the short call and the four background tools serve the long simulation, and the two kinds do not overlap, which meets \emph{orthogonality}. Seven tools with a few parameters each carry the whole workflow, which meets \emph{simplicity}.

\subsection{Documentation tools: agentic query and browse}
\label{sec:docs}

\begin{figure}[pos=t]
  \centering
  \includegraphics{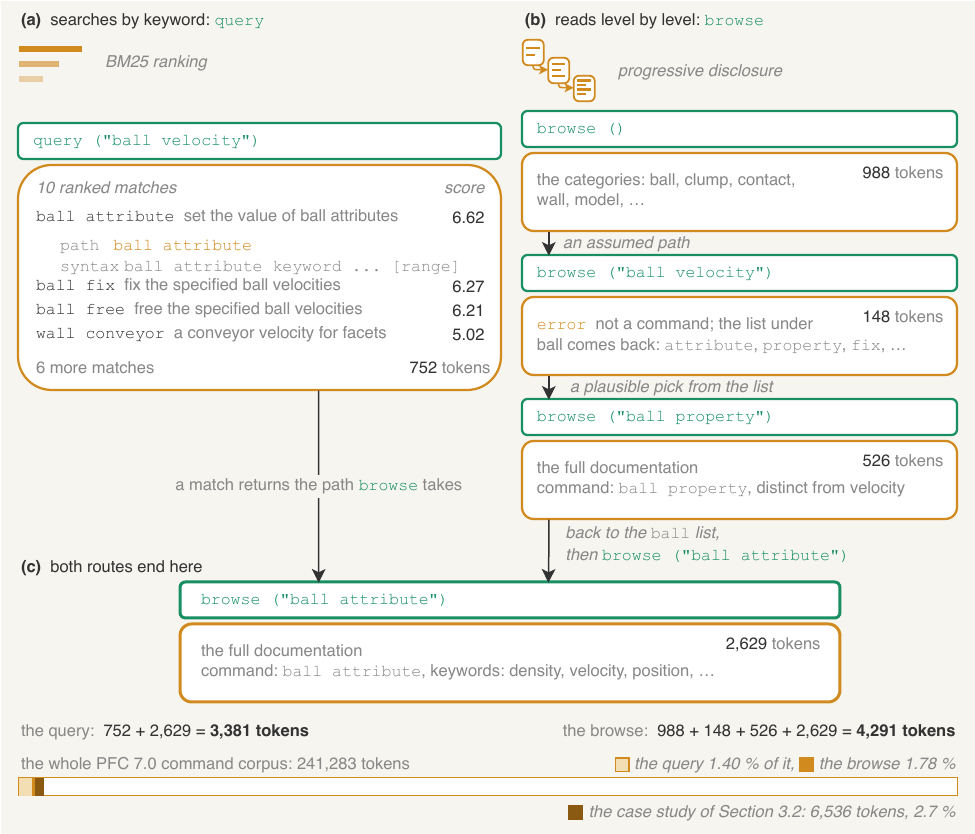}
  \caption{Two routes to one entry of the documentation, on the PFC 7.0 command corpus. Each response is counted in tokens. (a) The query tool searches by keyword and returns ranked matches, each with the path of the command. (b) The browse tool reads level by level, and a wrong path returns the list at the parent level. (c) Both routes end in the same browse call, which returns the full documentation. The bar shows the shares of the command corpus read by the documentation tools in examples (a) and (b) and in the case study of \cref{sec:campaign}.}
  \label{fig:docsearch}
\end{figure}

\cref{sec:interface} established a loop: the documentation guides the code the agent writes, and an error from the engine sends the agent back to the documentation. The query tool searches by keyword, with no embedding model and no vector database. Keyword search has been found sufficient in agent-driven search \citep{piserini2026}. The corpus is built from the engine's official documentation and verified in a running engine. The rest of this subsection describes the two tools in turn, each with an illustrative example in \cref{fig:docsearch}.

\textbf{The query tool.} The query tool searches the documentation by keyword. It takes one or more keywords, ranks the commands that match, and returns them. Each command in the corpus has three searchable fields: its name, a list of keywords, and its description. The ranking is BM25 \citep{robertson2009bm25}. The rarer a term is across the corpus, the more it counts in the ranking score. A keyword such as \texttt{conveyor}, which few commands use, therefore brings those few commands to the top, while a word such as \texttt{set}, which appears everywhere, adds little. A match in the name counts most and a match in the description counts least, because in a command corpus the name is the most informative field. Each match returns the path, the syntax, and a one-line description of the command. In \cref{fig:docsearch}(a) the agent wants to set the velocity of one ball in PFC and searches for the command by keyword. The query \texttt{ball velocity} returns ten matches: \texttt{ball attribute} first, \texttt{ball fix} and \texttt{ball free} next, because their descriptions are about velocities. The first match, described as setting the value of ball attributes, is plainly the one wanted, and the agent browses its path. The browse call returns the full documentation, and the lookup takes 3{,}381 tokens in total. The corpus is 241{,}283 tokens at this version, and the lookup read 1.40 percent of it.

\textbf{The browse tool.} The browse tool reads the documentation level by level. With no path, the tool lists the categories of the corpus (\texttt{ball}, \texttt{clump}, \texttt{contact}, \texttt{wall}, \texttt{model} and the rest). With the path of a category, the tool lists the commands in the category, each with the syntax and a one-line description. With the path of a command, the tool returns the full documentation of the command: the description, the keywords the command accepts, the examples, and the notes. This progressive disclosure \citep{arag2026} saves the context window, because each call returns only the level the agent asks for. A wrong path returns the list at the parent level, which drives the agent to browse the correct path. In \cref{fig:docsearch}(b) the agent looks for the same PFC command by reading level by level, and starts from the categories. It assumes the path \texttt{ball velocity}. That is not a command, and the error returns the names of the twenty commands under \texttt{ball}. Judging from the names in the list, the agent makes a plausible pick, \texttt{ball property}, but its documentation is about the properties of contact models, not velocity, and the agent goes back to the list. The third call is \texttt{ball attribute}, the command the agent wanted, and the lookup takes 4{,}291 tokens in total. The lookup read 1.78 percent of the corpus.

\Cref{fig:docsearch} shows each tool on its own, and both examples end in the same browse call. In practice the agent mixes the two tools: a wrong browse is followed by a query, and an unclear list of matches is followed by a browse of a category. The examples use the command language of PFC. The Python API, the FISH language and the reference material of PFC, and the Python API of YADE, are searched and read with the same query and browse tools.

\textbf{Summary.} The corpus is built from the engine's official documentation and verified in a running engine, which meets \emph{correctness}. The browse tool reaches every command in the documentation, which meets \emph{coverage}. The query tool searches the documentation and the browse tool reads the documentation, and the two tools do not overlap, which meets \emph{orthogonality}. The two tools need no embedding model and no vector database, which meets \emph{simplicity}.

\subsection{Execution tools: the threading model of the bridge in PFC}
\label{sec:bridge}

\begin{figure}[pos=t]
  \centering
  \includegraphics{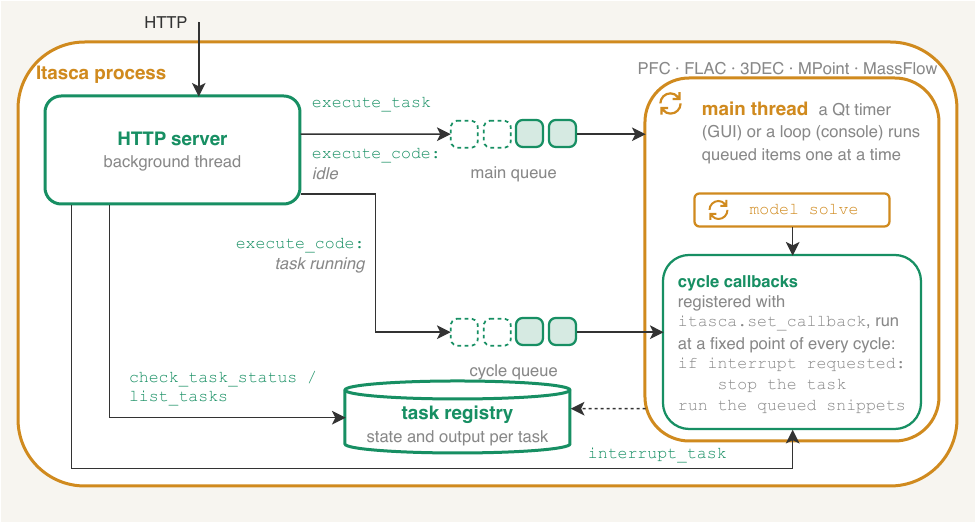}
  \caption{The threading model of the bridge in PFC. The HTTP server receives each request on a background thread. A script, and a snippet while no task is running, go into the main queue, and a Qt timer or a console loop runs the items of the main queue on the main thread. A snippet that arrives while a task is running goes into the cycle queue, and the snippet callback runs the snippet. The interrupt callback stops the task when an interrupt is requested. Status queries are answered from the task registry without a queue. Green: added by the bridge; amber: the engine.}
  \label{fig:itasca}
\end{figure}

\cref{sec:interface} designed the five execution tools in the MCP server box of \cref{fig:architecture}, and \cref{sec:docs} described the two documentation tools beside them. Unlike the documentation tools, which the MCP server answers from the corpora it carries, the execution tools have to reach the engine. The MCP server forwards each call of an execution tool to the engine, so the execution tools need a receiver inside the engine process. The receiver takes the agent's code, a snippet (from \executecode{}) or a script (from \executetask{}), runs the code in the engine, and returns the answer to the MCP server. That receiver is the bridge, the box inside the engine in \cref{fig:architecture}. The bridge is a small HTTP server inside the engine, with one address per execution tool. The split between the MCP server and the bridge has two consequences:
\begin{itemize}
  \item The bridge depends on nothing but Python's standard library, so nothing has to be installed into the Python each engine embeds, and the engine gains no dependency, which is what \emph{simplicity} asks of it.
  \item The MCP server runs with the agent and the bridge with the engine, so the agent's process is not tied to the engine's: the modeler can stop an agent and start another, and the engine and the model in it stay alive, which matters because a simulation takes hours or days.
\end{itemize}
\Cref{sec:interface} set three requirements for the execution tools: a script runs in the background as a task, a running task can be interrupted, and a snippet can be executed while a task is running. The threading model of the bridge is designed to meet these three requirements inside one engine process. The HTTP server of the bridge runs on a background thread, so that the server receives every request, whether the engine is idle or running a task (\cref{fig:itasca}). This subsection describes the challenges of serving the five tools inside PFC and how the bridge addresses them under the four principles of \cref{sec:interface}. \Cref{sec:semantics} describes the threading model of the bridge in YADE.

\textbf{Through the main queue.} Every script, and a snippet that arrives while no task is running, goes through the main queue (\cref{fig:itasca}, upper branch). Two constraints determine this route. First, a script or a snippet that contains engine commands must run on the main thread, the thread of the console. Second, the agent may submit a script while an earlier task is still running, and the tasks must run one at a time in the order submitted. The HTTP thread therefore places the code in the main queue, and the main thread takes from the main queue one item at a time. With the GUI open, a Qt timer takes the next item between two GUI events and runs it on the main thread; in console mode, a loop on the main thread takes and runs the items the same way. A task typically starts a solve and then computes for hours on the main thread. PFC refreshes its GUI during the solve, so the modeler keeps watching the DEM model. When a task finishes, the Qt timer or the loop takes the next script or snippet from the main queue, and continues until the main queue is empty.

\textbf{Through the task registry.} The status tools, \checktaskstatus{} and \listtasks{}, do not go through a queue. The HTTP thread answers them directly from the task registry of the bridge (\cref{fig:itasca}, bottom), which holds the identifier of each task, its state, and its output so far. A call of \checktaskstatus{} or \listtasks{} returns at once, even while a task occupies the main thread. The task history survives a restart of PFC.

\textbf{Through the cycle callbacks.} While a task is running, the Qt timer or the loop cannot take the next item from the main queue. \interrupttask{} and \executecode{} therefore need another route to the main thread. PFC offers a route into the running solve: \texttt{itasca.set\_callback} registers a function that the engine calls once in every cycle of a solve, on the main thread. The bridge registers two such callbacks, the interrupt callback and the snippet callback. Because \texttt{model new} and \texttt{model restore} clear the engine's callbacks, and a task script issues them freely, the bridge registers its two callbacks again before every engine command it issues.

\cref{sec:interface} required \interrupttask{} to stop a running task. The bridge records the interrupt request (\cref{fig:itasca}, bottom edge), and the interrupt callback checks for a request once in every cycle (\cref{fig:itasca}, bottom right). When the interrupt callback finds a request, the interrupt callback raises an exception inside the solve, which ends the task. The interrupt callback runs only while the engine cycles, but a task is not always cycling: it may be building the model, or may contain no solve at all. The bridge therefore runs the same check at every call of \texttt{itasca.command()} in the task script, so a requested interrupt stops the task at the next command.

The agent cannot write into the script every quantity it will later want to see, so \cref{sec:interface} required \executecode{} to stay available while a task runs. While a task runs, a snippet goes to a second queue, the cycle queue (\cref{fig:itasca}, middle branch), and the snippet callback runs the snippet in the next cycle, still on the main thread. The snippet runs within one step of the simulation, never across two, so all the values the snippet reads come from the same step. The task does not notice the pause. A snippet that contains a solve is refused, because the snippet callback is itself driven by the solve.

\textbf{Summary.} The threading model of the bridge in PFC meets the three requirements of \cref{sec:interface}. A script runs in the background as a task through the main queue. A running task is interrupted through the interrupt callback. A snippet is executed while a task is running through the cycle queue and the snippet callback. A snippet runs on the main thread and within one step of the simulation, which meets \emph{correctness}. The long simulation and the short call do not interfere with each other: a task never blocks a status query, an interrupt, or a snippet, which meets \emph{orthogonality}. The bridge consists of the Python standard library, one HTTP server, two queues, and two callbacks of the engine, which meets \emph{simplicity}.

\subsection{Yet another engine: the threading model of the bridge in YADE}
\label{sec:semantics}

\begin{figure}[pos=t]
  \centering
  \includegraphics{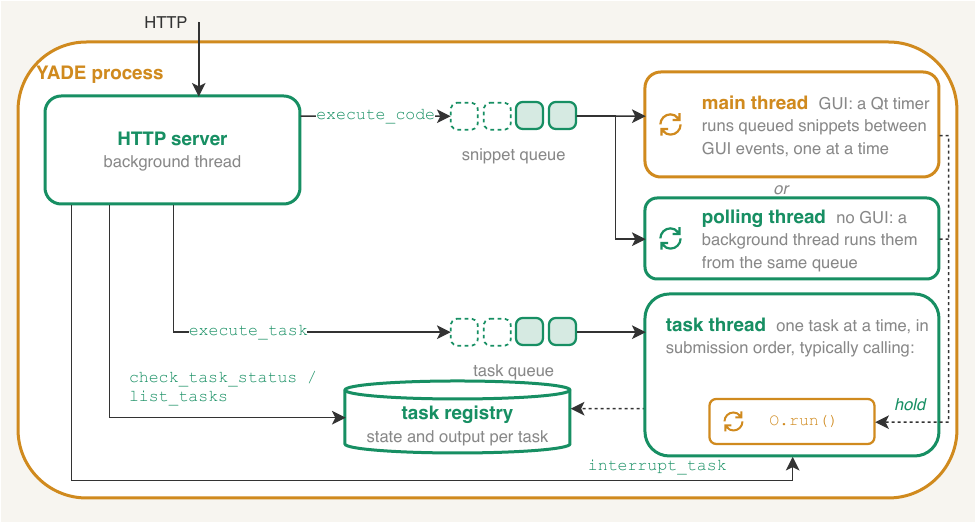}
  \caption{The threading model of the bridge in YADE. The HTTP server and the task registry are as in \cref{fig:itasca}. A YADE simulation can run on a background thread, and the bridge uses separate queues and separate threads for snippets and for tasks. A snippet goes into the snippet queue, and a Qt timer on the main thread runs the snippet, or a polling thread when Qt is not available. A script goes into the task queue, and the task thread runs the tasks in the order submitted. The dashed \emph{hold} edge lets a snippet hold the cycling simulation for a read (\cref{fig:runwrapper}). Green: added by the bridge; amber: the engine.}
  \label{fig:threads}
\end{figure}

\cref{sec:bridge} completed the path from the agent to a commercial engine: the tools of \cref{sec:interface}, the MCP server that forwards them, and the bridge inside PFC that serves them. YADE, the open-source engine, has the same path, and its bridge is of a similar construction (\cref{fig:threads}): the same HTTP server, the same five addresses, a snippet queue and a task queue, and the same registry. YADE differs from PFC in the threads on which the agent's code may run, and the threading model of the bridge in YADE differs accordingly.

Three facts about YADE decide the design:
\begin{itemize}
  \item The YADE engine can be called from the main thread and from a background thread.
  \item Its GUI, built on Qt, runs on the main thread only.
  \item A task script starts a simulation and waits until the simulation finishes, and this wait on the main thread would stop the GUI from refreshing.
\end{itemize}
The bridge therefore runs the code snippet (from \executecode{}) and the task script (from \executetask{}) on two different threads (\cref{fig:threads}). The code snippet goes into the snippet queue. When Qt is available in the YADE process, a Qt timer on the main thread takes the snippets from the snippet queue in order and runs each snippet between two GUI events. The agent can therefore use the GUI from a snippet: open a view of the DEM model, or capture the view as an image for the agent's own inspection or for the modeler. When Qt is not available, a loop on a polling thread takes the snippets from the snippet queue in order and runs each snippet. The task script goes into the task queue. A loop on the task thread, a background thread, runs the scripts in the task queue in order. The main thread therefore stays free, and the GUI keeps refreshing while a task computes. As a result, during a long task the modeler and the agent can observe and control the GUI. The rest of this subsection describes how the bridge interrupts a running task, and how a snippet reads the DEM model while the task thread is cycling the simulation.

\begin{figure}[pos=htbp]
  \centering
  \includegraphics{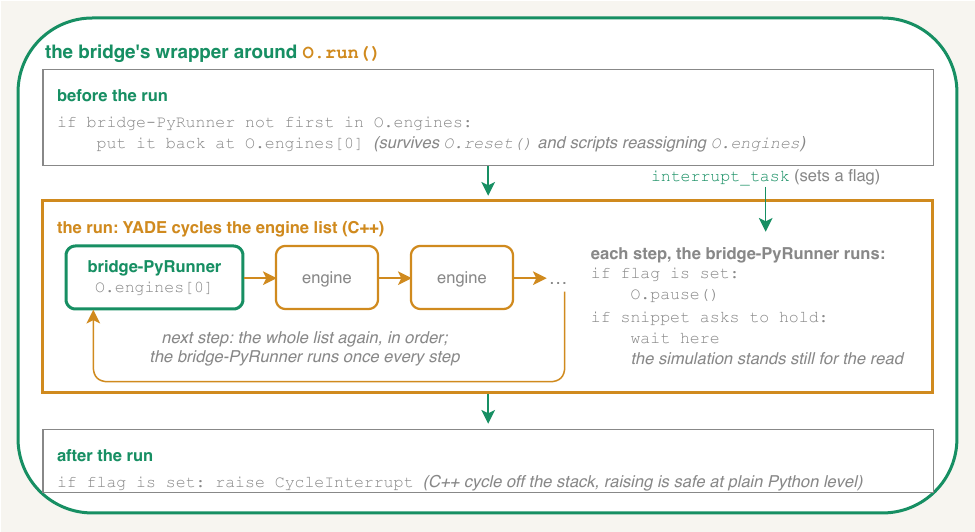}
  \caption{The bridge's wrapper around \texttt{O.run()}, a zoom into the \texttt{O.run()} box of \cref{fig:threads}. The wrapper wraps every \texttt{O.run()} call, typically the one a task script starts on the task thread. Before the simulation, the wrapper restores the bridge-PyRunner to the head of the engine list. During the simulation, the bridge-PyRunner runs once in every step. The bridge-PyRunner calls \texttt{O.pause()} when an interrupt is requested, and holds the simulation while a snippet reads the DEM model. After the simulation, outside the cycle, the wrapper raises \texttt{CycleInterrupt} if an interrupt was requested. Green: added by the bridge; amber: the engine's existing cycle.}
  \label{fig:runwrapper}
\end{figure}

\textbf{Interrupting a task.} As in PFC, the agent interrupts a task gone wrong (\cref{sec:interface}). YADE provides a counterpart of the callback in PFC: a PyRunner, an engine that returns control to Python at prescribed steps. The bridge registers one, the bridge-PyRunner, at the head of the engine list, so that it runs a short check at the head of every step (\cref{fig:runwrapper}). A call of \interrupttask{} sets a flag. When the bridge-PyRunner finds the flag, the bridge-PyRunner calls \texttt{O.pause()} rather than raising an exception, because an exception raised inside YADE's cycle ends in a C++ fatal error. The simulation stops at the end of the current step, and \texttt{O.run()} returns normally. Back in Python, outside the \texttt{O.run()} cycle, the bridge raises \texttt{CycleInterrupt}, which ends the task (\cref{fig:runwrapper}).

A script can remove or displace the bridge-PyRunner: \texttt{O.reset()} clears the engine list, and an assignment to \texttt{O.engines} replaces the engine list. The bridge therefore wraps \texttt{O.run()}. Before every call of \texttt{O.run()}, the wrapper restores the bridge-PyRunner to the head of the engine list (\cref{fig:runwrapper}).

\textbf{Holding the cycle for one step.} Running the code snippet and the task script on two threads has a cost. In PFC the snippet callback runs the snippet at a fixed point of a cycle, on the main thread. In YADE a snippet that arrives while a task is running goes into the snippet queue and runs on the main thread, because a snippet can use the GUI only from the main thread. The main thread and the task thread run concurrently: the task thread keeps cycling the simulation while the snippet runs on the main thread. The situation is like a photograph taken from a fast-moving train, which comes out distorted by the rolling-shutter effect: the sensor reads out line by line while the train moves, so each line captures a different place. A snippet reading a cycling simulation suffers the same distortion across time steps: two values it reads can come from two different steps, which violates \emph{correctness}. The requirement is therefore that the snippet completes within one step, never across two. The snippet asks the running cycle to hold. The simulation waits at the bridge-PyRunner, and the snippet reads the model while the task stands still. The hold is not an interrupt: the moment the snippet finishes, the task continues. The bridge refuses a snippet that cycles the simulation while a task is cycling, as in PFC.

\textbf{Summary.} The threading model of the bridge in YADE meets the three requirements of \cref{sec:interface} through the same five tools as in PFC. An interrupt stops a task while the YADE process keeps running. A snippet runs within one step of the simulation, which meets \emph{correctness}. A task runs on its own thread and never blocks a status query, an interrupt, or a snippet, which meets \emph{orthogonality}. The bridge adds to YADE one engine at the head of the engine list and one wrapper around \texttt{O.run()}, which meets \emph{simplicity}.

\subsection{Beyond the discrete element method}
\label{sec:portability}

The interface is not limited to the discrete element method. The bridge of \cref{sec:bridge} serves, with the same tools and the same construction, the rest of the Itasca suite: continuum analysis (FLAC), distinct-element blocky rock (3DEC), the material point method (MPoint) and gravity flow of caved rock in mines (MassFlow). The engines above advance the simulation in small time steps. Each step takes little time to compute, and the state of the simulation is consistent at the end of every step. The interrupt and the snippet of \cref{sec:bridge,sec:semantics} act between two steps, and the engine reaches such a point frequently. The tool design therefore suits engines with explicit time stepping, commercial or open-source.
 \section{Case study: agent-driven calibration of a drained triaxial model}
\label{sec:casestudy}

\Cref{sec:server} designed the tools through which an agent operates a DEM engine and the bridge that serves the tool requests inside the engine. This section connects a general-purpose agent to the engine through those tools. The task assigned to the agent is to calibrate a drained triaxial model against laboratory data, in a live engine. The task is chosen for two reasons. First, the triaxial test is a standard laboratory test in geotechnical engineering, and calibrating a DEM model against a triaxial test is a routine step of DEM modeling. Second, the task tests two abilities of the agent: writing the commands of the engine correctly, and relating the parameters of the DEM model to the stress--strain response of the specimen. The modeler provides the agent with the laboratory curves and instructs the agent to calibrate the DEM model against the curves. \Cref{sec:protocol} sets the calibration task. \Cref{sec:campaign} reports the session in PFC. \Cref{sec:yade} gives the same task to the agent in YADE. \Cref{sec:models} gives the same task to seven language models. \Cref{sec:pipeline} reports a smaller task, given to a small language model in the test pipeline of YADE.

\subsection{Setting the task}
\label{sec:protocol}

The calibration starts from a laboratory test, a drained triaxial compression test on dense Toyoura sand by \citet{fukushima1984}. The specimen was prepared to a relative density of about 82\,\%, isotropically consolidated to $\sigma_3' = 98$ kPa (1.0 kgf/cm$^2$ in the source), and sheared to 13.5\,\% axial strain. A calibration is iterative: the parameters of a DEM model are adjusted and the test is simulated repeatedly, until the simulated test reaches the targets, a few quantities taken from the laboratory test. This subsection specifies the constants of the DEM model (the specimen, the contact law, and the test procedure), the parameters to be calibrated, and the targets.

\textbf{The constants of the DEM model.} The DEM model is an assembly of spheres with a rolling-resistance linear contact law \citep{belheine2009}, in a cube between six rigid frictionless walls. The particles are 5000 spheres with diameters uniformly distributed between 1 and 3 mm, the grading of Toyoura sand at ten times the size. The stiffness ratio is fixed at the value of an earlier model of the same sand \citep{han2026hca}, and the local damping \citep{cundall1987} at the value usual for quasi-static simulations \citep{marketos2010}. The simulation of the test has three steps. In preparation, the spheres are generated as a loose packing and compressed by the walls, without friction, to a prescribed porosity. In consolidation, the contacts are given their stiffness and friction, and the specimen is consolidated isotropically to 98 kPa. In shear, the specimen is sheared at constant axial strain rate and constant lateral stress until past the peak deviator stress. An iteration is one set of parameters and the simulation of the test with that set. \Cref{tab:spec} lists the constants of the DEM model and of the three steps.

\textbf{The calibrated parameters.} Four parameters are calibrated: the contact modulus $E^*$, the sliding friction $\mu$ and the rolling friction $\mu_r$ of the contact law, and the porosity $n_0$ of the frictionless specimen at the end of preparation. The porosity is calibrated because the density of the specimen acts on dilatancy and peak strength. The ranges in \cref{tab:spec} are the usual search ranges of these parameters.

\begin{table}[pos=htbp]
\caption{Model specification. The four parameters to calibrate, with their ranges, and the constants.}
\label{tab:spec}
\begin{tabular}{@{}lp{0.72\linewidth}@{}}
\toprule
\multicolumn{2}{@{}l}{\emph{Calibrated}} \\
\midrule
$n_0$ & porosity of the frictionless specimen, before consolidation, 0.380--0.445 \\
$E^*$ & contact modulus, $10^7$--$10^9$ Pa \\
$\mu$ & sliding friction, 0--1 \\
$\mu_r$ & rolling friction, 0--1 \\
\midrule
\multicolumn{2}{@{}l}{\emph{Constant in every session}} \\
\midrule
contact model & rolling-resistance linear, $k_n/k_s$ = 2.0 \\
particles & 5000 spheres, diameter uniform in 1--3 mm, radii drawn once from a fixed seed, $\rho$ = 2640 kg/m$^3$ \\
specimen & cube between six rigid frictionless walls \\
preparation & generated loose, compressed by the walls at $\mu$ = $\mu_r$ = 0 to $n_0$ \\
consolidation & at the declared $\mu$ and $\mu_r$; isotropic, all six walls servo-controlled to 98 kPa \\
shear & axial strain rate 0.5 s$^{-1}$; lateral walls servo-controlled to 98 kPa; ends 1\,\% axial strain past the peak deviator stress once $q$ has fallen 3\,\% below it, or at 15\,\% \\
damping & local, 0.7 \\
budget & 12 task submissions, 8 h of session time \\
\bottomrule
\end{tabular}
\end{table}

\textbf{The targets.} The calibration is scored on three target scalars (\cref{tab:targets}): the secant modulus $E_{50}$ at half the peak deviator stress for stiffness, the peak deviator stress $q_\mathrm{peak}$ for strength, and the dilatancy rate $D = \mathrm{d}\varepsilon_v/\mathrm{d}\varepsilon_a$ at the peak deviator stress for dilatancy, the quantities a DEM calibration of a triaxial test conventionally targets \citep{belheine2009,coetzee2017}. Each target has a tolerance. The calibration task succeeds when all three simulated values are within their tolerances in the same iteration. The tolerances differ with the accuracy of the laboratory values. $q_\mathrm{peak}$ is tabulated in the source and has the tightest tolerance, 10\,\%. $E_{50}$ also needs the axial strain at half the peak deviator stress, which is read from the steep initial part of a digitized curve, where a small error in strain is a large error in modulus, and has 25\,\%. $D$ is a slope and has 0.10 in absolute terms. The post-peak part of the curve is not a target, because strain localization in a dense specimen contaminates it. The critical state is not a target, because the test ends with the stress ratio still falling and the volume still growing; dense sand at this pressure does not reach the critical state within 15\,\% axial strain \citep{VerdugoIshihara1996}.

Each error $e_i$, with $i$ = 1 for $E_{50}$, 2 for $q_\mathrm{peak}$ and 3 for $D$ (\cref{tab:targets}), is divided by its tolerance $\tau_i$ and keeps its sign, so that $e_i/\tau_i$ within $\pm 1$ is within tolerance and the direction of the miss is visible. The largest of the three magnitudes, $J = \max_i |e_i|/\tau_i$, summarizes an iteration in one number. The criterion of success is therefore $J \le 1$, and the agent stops at the first iteration that reaches $J \le 1$. The budget is 12 task submissions and 8 h of session time. If the budget runs out first, the iteration with the lowest $J$ is reported.

\begin{table}[pos=htbp]
\caption{Calibration targets and tolerances. The error $e_i$ is the simulated value minus the target, relative to the target or absolute as marked; an iteration is scored by $J = \max_i |e_i|/\tau_i$, and every target is within tolerance when $J \le 1$.}
\label{tab:targets}
\begin{tabular}{@{}rlllll@{}}
\toprule
$i$ & Scalar & Definition & Target & $e_i$ & $\tau_i$ \\
\midrule
1 & $E_{50}$ & secant modulus at half the peak deviator stress & 52.5 MPa & relative & 25\,\% \\
2 & $q_\mathrm{peak}$ & peak deviator stress & 403 kPa & relative & 10\,\% \\
3 & $D$ & $\mathrm{d}\varepsilon_v/\mathrm{d}\varepsilon_a$ at the peak deviator stress & 0.67 & absolute & 0.10 \\
\bottomrule
\end{tabular}
\end{table}

\subsection{Calibration in the live engine}
\label{sec:campaign}

\begin{figure}[pos=t]
  \centering
  \includegraphics{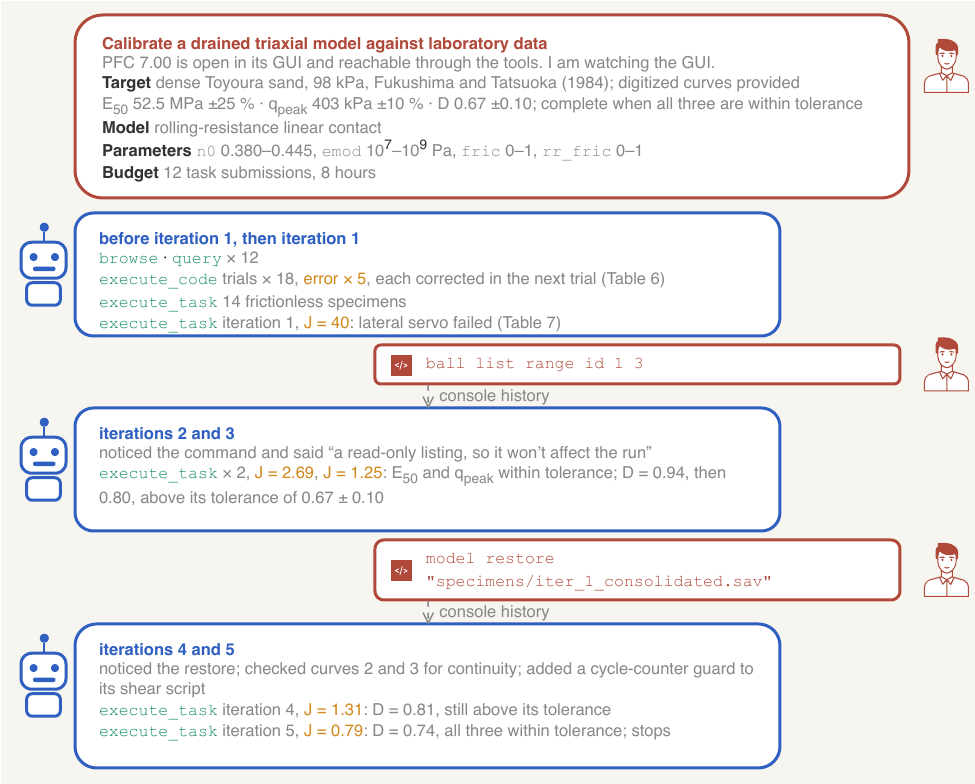}  \caption{The session as a conversation, read from top to bottom: the modeler's opening message, abridged (the full text is published, see \hyperref[sec:data]{Data availability}); the agent's messages on the left, grouped by iteration and summarized by their tool calls and outcomes; and the modeler's two console actions on the right, which reached the agent as console history.}
  \label{fig:timeline}
\end{figure}

With the task set, this subsection reports the calibration. The session is one conversation between the modeler and the agent, and one PFC 7.0 process on a Windows workstation with its graphical interface open. The agent is a general-purpose agent and reaches the engine only through the tools of \cref{sec:server}. The versions of the agent and the tools are listed under \hyperref[sec:data]{Data availability}. The task of \cref{sec:protocol} is written into the message that opens the session, the first message of \cref{fig:timeline}. During the session every tool call of the agent is approved automatically and runs at once. After the opening message the modeler says nothing in the conversation. The modeler types two commands at the console of the engine, both planned before the session, to test whether the agent notices a command typed at the console and how the agent judges the effect of the command on its simulation. The opening message, the agent's messages and tool calls, the engine's output for every task and the commands typed at the console are all recorded and published with the test data (\hyperref[sec:data]{Data availability}).

\textbf{First simulation.} In the first ten minutes the agent looked up the contact model, the wall servo, the contact stiffness method, the wall generator, the range syntax, the timestep command and the cycle callback in the documentation, with twelve calls of the query and browse tools of \cref{sec:docs}. The twelve calls returned 12{,}882 tokens together; the seven that went to the command corpus of \cref{fig:docsearch} returned 6{,}536 tokens, 2.7\,\% of that corpus.

Between the lookups, and without being told to, the agent ran eighteen small trials in the engine through \executecode{}, most of them on models of one or two particles. The trials checked the syntax, that is, whether the engine accepted a command or a call. The trials also checked in numbers what the commands did: the stiffness of a contact for a given contact modulus, and the speed at which the engine cycled with 5000 particles. Five of the eighteen trials came back with an error from the engine, three in the Python API of PFC and two in the command language of PFC (\cref{tab:trialerrors}). The agent ran four of the five without looking in the documentation first; in the fifth, the agent left out a command that its earlier trials had contained. Each was corrected in the next trial, from what the engine returned.

\begin{table}[pos=htbp]
\caption{The five trials that came back with an error from the engine.}
\label{tab:trialerrors}  \small
\begin{tabular}{@{}>{\raggedright\arraybackslash}p{0.30\linewidth}>{\raggedright\arraybackslash}p{0.27\linewidth}>{\raggedright\arraybackslash}p{0.36\linewidth}@{}}
\toprule
The agent ran & Cause & Solution \\
\midrule
\texttt{w.velocity()} & A wall has no such attribute. & Listed the methods of the wall through \executecode{}; used \texttt{w.vel()}. \\
\addlinespace
\texttt{ball.list()[0]} & The list of balls cannot be indexed. & Removed the index. \\
\addlinespace
\texttt{itasca.cycle(20000)} & The function returns the cycle number and does not cycle. & Used the command \texttt{model cycle 20000}. \\
\addlinespace
\texttt{program list threads} & \texttt{threads} is not a keyword of \texttt{program list}. & Used \texttt{program list information}, from the keywords listed in the error. \\
\addlinespace
\texttt{model cycle 1} & \texttt{model large-strain on} was left out. & Added the command, named in the error. \\
\bottomrule
\end{tabular}
\end{table}

The agent then wrote two scripts into its working directory: one that prepares frictionless specimens and one that consolidates and shears a specimen at a given parameter set. Its first submission through \executetask{} prepared not one specimen but fourteen, one at every porosity from 0.445 to 0.380 in steps of 0.005, in one task; every later iteration took its specimen from this set. The agent declared its first set, iteration 1 in \cref{tab:trajectory}, $n_0$ = 0.385, $E^*$ = $10^8$ Pa, $\mu$ = 0.5, $\mu_r$ = 0.25, with its reasoning: the peak stress ratio and dilatancy rate of the test, through Rowe's stress--dilatancy relation \citep{rowe1962}, ask for a sliding friction near 0.5 with moderate rolling resistance in a dense packing, and a secant modulus taken at half the peak asks for a contact modulus of about twice $E_{50}$. The agent submitted the script through \executetask{} and polled for the simulation to finish. The modeler watched the same simulation in the graphical interface.

The first simulation came back with a result that made no physical sense: a peak strength five times the target and a negative dilatancy rate. The agent did not attribute the result to the parameters. It read the lateral stress column its script had recorded and found that the lateral walls had reached the velocity cap of the servo at 1\,\% axial strain: the cap was too low for the rate at which the dilating specimen pushed the walls outward, and the lateral stress had then climbed to six times its target. The agent declared the simulation invalid.

\begin{table}[pos=htbp]
\caption{Parameters and scores of each iteration. $\varepsilon_{a,\mathrm{peak}}$ is the axial strain at the peak deviator stress; $e_1$, $e_2$ and $e_3$ are the errors of $E_{50}$, $q_\mathrm{peak}$ and $D$, each divided by its tolerance $\tau_i$ (\cref{tab:targets}). Iteration 1 is the simulation whose lateral walls reached their velocity cap; its row is kept because the submission was spent.}
\label{tab:trajectory}  \small
\begin{tabular}{@{}rrrrrrrrrr@{}}
\toprule
it. & $n_0$ & $E^*$ (Pa) & $\mu$ & $\mu_r$ & $\varepsilon_{a,\mathrm{peak}}$ & $e_1/\tau_1$ & $e_2/\tau_2$ & $e_3/\tau_3$ & $J$ \\
\midrule
1 & 0.385 & $1.0 \times 10^8$ & 0.50 & 0.25 & \multicolumn{5}{c}{invalid, lateral stress off target} \\
2 & 0.385 & $1.8 \times 10^8$ & 0.50 & 0.25 & 2.9\,\% & $+0.44$ & $-0.97$ & $+2.69$ & 2.69 \\
3 & 0.395 & $1.8 \times 10^8$ & 0.50 & 0.50 & 3.0\,\% & $-0.29$ & $+0.63$ & $+1.25$ & 1.25 \\
4 & 0.400 & $2.0 \times 10^8$ & 0.50 & 0.50 & 3.4\,\% & $-0.23$ & $+0.18$ & $+1.31$ & 1.31 \\
5 & 0.410 & $2.2 \times 10^8$ & 0.50 & 0.70 & 5.5\,\% & $-0.79$ & $+0.21$ & $+0.61$ & 0.79 \\
\bottomrule
\end{tabular}
\end{table}

\begin{figure}[pos=t]
  \centering
  \includegraphics{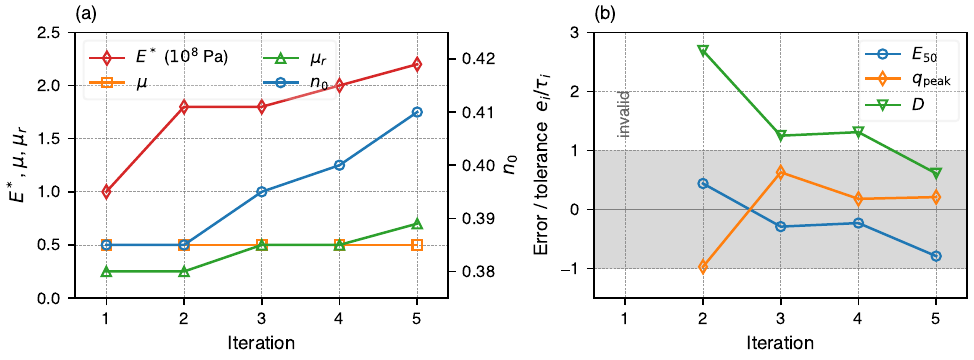}  \caption{Convergence of the session. (a) The four parameters declared for each iteration. (b) Signed errors of the three targets, each divided by its tolerance, against iteration; the shaded band is the tolerance, and iteration 1 has no score because its lateral walls reached their velocity cap.}
  \label{fig:conv}
\end{figure}

\textbf{Iterations 2 and 3.} Iteration 2 kept the porosity $n_0$, raised the servo cap, and changed one parameter, $E^*$ from $10^8$ to $1.8 \times 10^8$ Pa (\cref{tab:trajectory}), to obtain the first valid measurement of strength and dilatancy at the friction the agent had chosen. The agent raised $E^*$ because the initial stiffness in iteration 1 was too low. The simulation began by restoring the frictionless specimen at 0.385, applied the new parameters, and consolidated. With the servo corrected, the lateral stress stayed within 1.1\,\% of 98 kPa. The simulation scored $J$ = 2.69: stiffness 11\,\% above target, strength 9.7\,\% below it, at the lower edge of its 10\,\% tolerance, and the dilatancy rate far outside its tolerance, 0.94 against 0.67. Iteration 3 doubled $\mu_r$ and loosened $n_0$ to 0.395 (\cref{tab:trajectory}); stiffness and strength came into their bands, and $D$ fell from 0.94 to 0.80 (\cref{fig:conv}). The strategy of the agent worked: the looser packing lowered the dilatancy, and the higher rolling friction raised the strength. The dilatancy rate was still above its tolerance.

\begin{figure}[pos=t]
  \centering
  \includegraphics{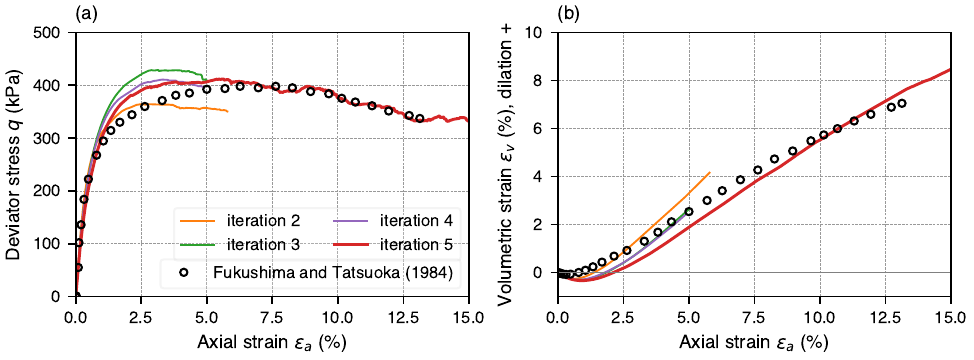}  \caption{(a) Deviator stress and (b) volumetric strain against axial strain: the test (points), iterations 2 to 4 (thin lines), and iteration 5, the calibrated set (thick line). Iterations 2 to 4 end past their peaks, as the message specified; the calibrated set is shown from a repetition of iteration 5 to 15\,\% axial strain, which reproduced the session's curve exactly up to the point where the session had stopped it.}
  \label{fig:curves}
\end{figure}

\textbf{Modeler and agent share one engine.} The agent shares the live engine with the modeler, and the MCP tools let the agent see what the modeler does there: every command typed at the engine's console returns to the agent as console history with its next tool result (\cref{fig:architecture}). The first of the two planned commands reads the model and changes nothing: during iteration 1 the modeler typed a listing of the particles with identifiers 1 to 3 at the command prompt (\cref{fig:timeline}). The engine ran the command while the task was in progress and recorded it, and the line reached the agent with its next tool call, the submission of iteration 2. The agent noted the line in one sentence, ``a read-only listing, so it won't affect the run'', and went on.

The second planned command replaces the model in the engine: between iterations 3 and 4 the modeler restored the consolidated state of iteration 1 at the command prompt. The agent saw the restore when it submitted iteration 4 and judged that the simulation was not affected, because its script begins by restoring its own specimen. The agent did not stop there. In case the modeler's action had reached beyond the model in memory, it went back over the recorded curves of iterations 2 and 3 and checked them for continuity, uniform strain steps and a monotonic cycle counter; both were intact, and it went on. The agent also added a guard to its shear script: in iteration 5, the guard checked during the simulation that the cycle counter had advanced by exactly the cycles the script ran. A restore from outside during the simulation would have aborted the simulation instead of corrupting its result. The moment is in the published record (\hyperref[sec:data]{Data availability}), in the agent's own notes under iteration 4. The modeler mentioned neither of the two commands in the conversation; the agent noticed both commands because the modeler ran the commands on the engine.

\textbf{Convergence.} Iteration 4 loosened $n_0$ by a further 0.005, with $E^*$ raised to offset the stiffness the looser packing would lose; $D$ did not move: the step had been too cautious. Iteration 5 took a larger step on both, $n_0$ to 0.410 and $\mu_r$ to 0.7; all three targets were reached at once, $J$ = 0.79, and the agent stopped, as the modeler's opening message required. The calibration took five iterations, of which iteration 1 was invalid (\cref{tab:trajectory}, \cref{fig:conv}). The peak had moved from 2.9\,\% axial strain in iteration 2 to 5.5\,\%, near the 6.3\,\% of the test (\cref{fig:curves}). The session used 6 of its 12 submissions, and the modeler intervened at no level: no message after the opening one, no denied call, no interrupt, no edit of a file; the session ran in one conversation and one engine process.

\subsection{Yet another engine}
\label{sec:yade}

\textbf{The same task on YADE.} \Cref{sec:server} described one set of MCP tools with two realizations, one inside the commercial DEM engine PFC and one inside the open-source DEM engine YADE, and the session above exercised the first. The agent was given the same task on YADE, in a second session, with the same agent and language model, and a message that differs from the first only in engine facts; the one that matters below is that YADE's rolling-resistance law writes the rolling friction as $\eta_r$ where PFC writes $\mu_r$. The targets, the tolerances, the definition of $n_0$, the procedure and the budget are unchanged.

\begin{figure}[pos=t]
  \centering
  \includegraphics{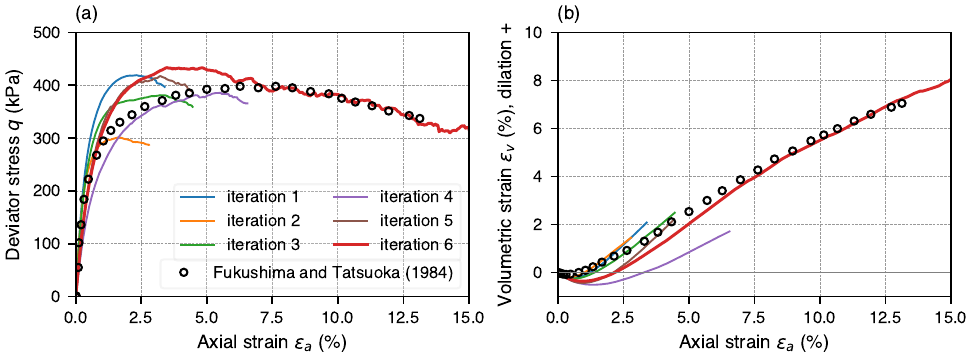}  \caption{(a) Deviator stress and (b) volumetric strain against axial strain on YADE: the test (points), iterations 1 to 5 (thin lines), and iteration 6, the calibrated set (thick line), shown from a repetition to 15\,\% axial strain that reproduced the session's curve exactly over its range.}
  \label{fig:curves-yade}
\end{figure}

\textbf{Calibrated in six iterations.} The agent looked up the contact material, its interaction functor, the contact law and the triaxial boundary controller in the documentation, six lookups in all, and tested the syntax it had read against the engine in twelve calls of \executecode{}. It wrote its scripts, submitted the preparation of the frictionless specimens as its first task, and ran the iterations in two more tasks of three parameter sets each. Iteration 6 reached all three targets, $J$ = 0.86, and the agent stopped (\cref{fig:curves-yade}). The session used 3 of its 12 submissions; the modeler typed nothing after the opening message.

The calibrated set on YADE is $n_0$ = 0.405, $E^*$ = $2.5 \times 10^8$ Pa, $\mu$ = 0.5 and $\eta_r$ = 0.35, with the peak at 3.5\,\% axial strain. The two engines write the same contact law with different conventions. For two spheres of radii $r_1$ and $r_2$ under a normal force $F_n$, PFC sets the normal stiffness $k_n = E^* \pi r_{\min}^2 / (r_1 + r_2)$ and the rolling limit $M^* = \mu_r F_n\, r_1 r_2 / (r_1 + r_2)$, whereas YADE sets $k_n = 2 E^* r_1 r_2 / (r_1 + r_2)$ and $M^* = \eta_r F_n\, r_{\min}$. For equal radii the same contact therefore needs $E^*$ on YADE $\pi/2 \approx 1.6$ times that on PFC, and $\eta_r = \mu_r / 2$. With these conventions, the calibrated set on YADE has the same sliding friction as the set on PFC, $\mu$ = 0.5, the same rolling friction, $\eta_r$ = 0.35 against $\mu_r$ = 0.7, and a lower modulus, $2.5 \times 10^8$ Pa against a translated $3.5 \times 10^8$ Pa (\cref{tab:trajectory}).

\begin{figure}[pos=t]
  \centering
  \includegraphics{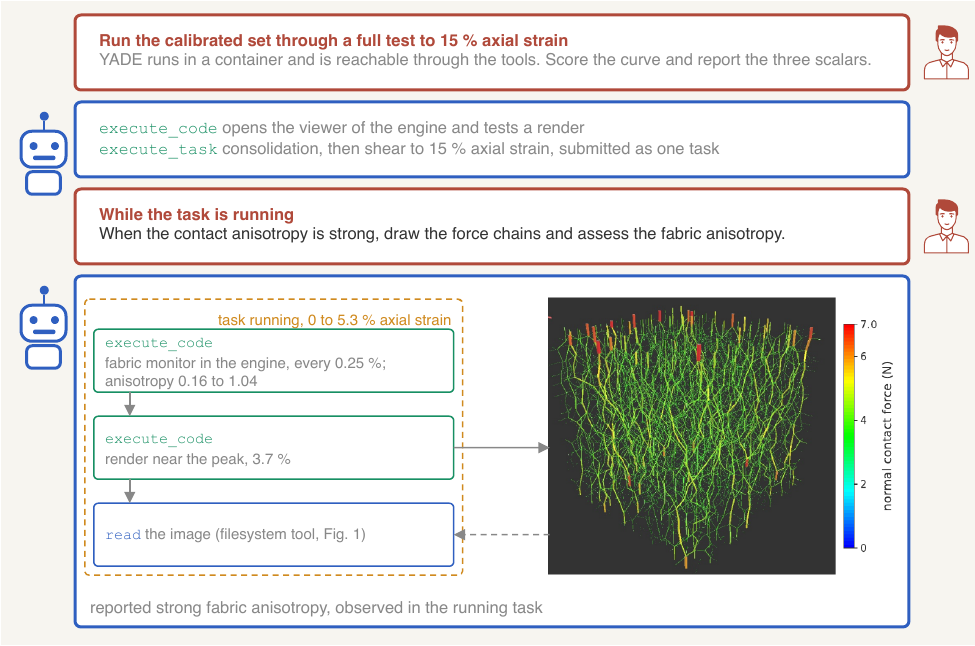}  \caption{The third session as a conversation, abridged, read from top to bottom: the modeler's opening message, the agent's submission of the task, the modeler's request while the task was running, and the agent's tool calls during the task, followed by the agent's report. The force chains were rendered by the engine's own viewer at 3.7\,\% axial strain, near the peak of the deviator stress. The task script contained neither the fabric monitor nor the render.}
  \label{fig:session3}
\end{figure}

\textbf{Reading a running task.} The MCP tools let the agent execute a code snippet while a task is running (\cref{sec:bridge,sec:semantics}). The two sessions above did not use this capability: every \executecode{} call came before the agent's first task. A third session on YADE used the capability (\cref{fig:session3}). The modeler asked the agent to run the calibrated set, iteration 6, to 15\,\% axial strain with the script of the earlier session. While the shear was running, the modeler asked in the conversation for the force chains and the fabric of the contacts, which the script did not record. The agent read the fabric of the running specimen through \executecode{}, set the running engine to record the contact anisotropy as the strain grew, rendered the force chains near the peak of the deviator stress, and read the image back through its filesystem tool (\cref{fig:architecture}); the curve this simulation recorded is identical to the repetition of iteration 6 in \cref{fig:curves-yade}, so the reads and the renders had not disturbed the physics. \Cref{sec:campaign} showed how the agent sees, through the MCP tools, what the modeler does in the engine during a running task. This session showed how the agent reads any variable of the model while a task is running, and even operates the viewer of the engine. Each session showed part of these capabilities, and both MCP servers, itasca-mcp \citep{itascamcp2026} and yade-mcp \citep{yademcp2026}, provide all of them: each offers the tools of \cref{tab:tools}, returns the console history with every tool result, and reads the model of a running task through \executecode{} (\cref{sec:bridge,sec:semantics}).

\subsection{The same task across language models}
\label{sec:models}

The two sessions above were run by one language model in one agent, and leave open whether other language models complete the task through the same tools. This subsection gives the task of \cref{sec:yade} to seven language models, three times each, and reports the results.

\begin{figure}[pos=t]
  \centering
  \includegraphics{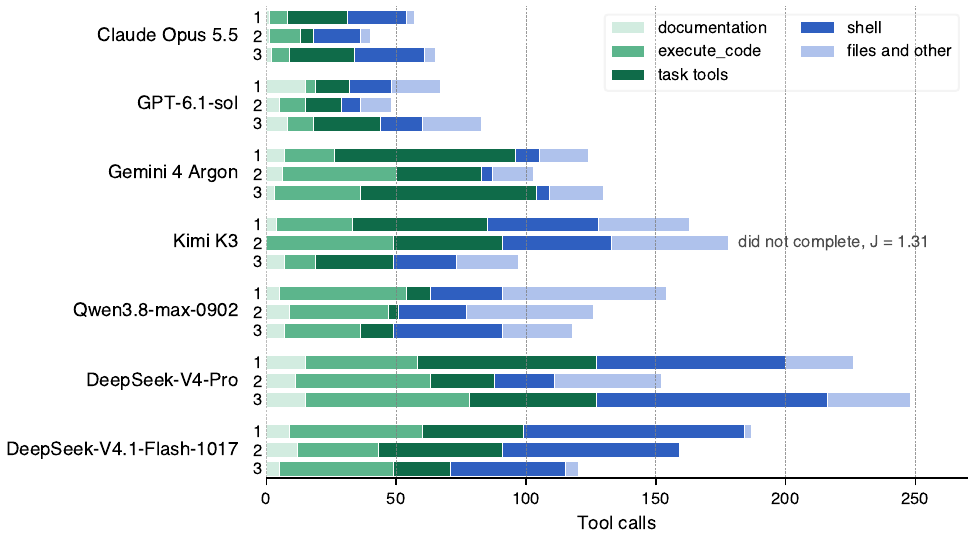}  \caption{Tool calls of each session, by tool: in green the MCP tools, the documentation tools, \executecode{} and the task tools, \executetask{} with \checktaskstatus{}, \interrupttask{} and \listtasks{}; in blue the agent's own tools, the shell and the rest. Sessions 1 to 3 of each language model; the session that did not complete is marked.}
  \label{fig:bench-tools}
\end{figure}

\textbf{One task, one set of MCP tools, seven language models.} To compare language models, everything but the language model is kept the same. The agent was the open-source OpenCode \citep{opencode2026}, with yade-mcp as its MCP server and its own filesystem and shell tools. The language models are Claude Opus 5.5, GPT-6.1-sol, Gemini 4 Argon, Kimi K3, Qwen3.8-max-0902, DeepSeek-V4-Pro and DeepSeek-V4.1-Flash-1017. Each session started in a fresh working directory holding the message of \cref{sec:yade} and the laboratory curve, and with a fresh YADE process. Every tool call was approved automatically and nobody intervened. The two language models from DeepSeek were reached through DeepSeek's API \citeyearpar{deepseekapi2026}, the other five through OpenRouter \citeyearpar{openrouter2026}, served by their vendors themselves. Each provider's default sampling temperature was used. Each session left an event log with every tool call, its result, its tokens and its cost, and the bridge log of the engine; \cref{tab:bench} and \cref{fig:bench-tools} are read from these (see \hyperref[sec:data]{Data availability}).

\begin{table}[pos=htbp]
\caption{Seven language models on the same task as in \cref{sec:yade}, three sessions each; each entry gives the lowest and highest of the three sessions. Completed: sessions that reached $J \le 1$; submissions: tasks submitted to the engine; context: tokens in the agent's context window at the end of the session; USD: cost.}
\label{tab:bench}
\footnotesize
\begin{tabular}{@{}lccccc@{}}
\toprule
model & completed & $J$ & submissions & context ($10^3$ tokens) & USD \\
\midrule
Claude Opus 5.5 & 3/3 & 0.59--0.95 & 3--4 & 85--85 & 1.69--1.99 \\
GPT-6.1-sol & 3/3 & 0.53--0.99 & 3--8 & 45--65 & 0.36--0.64 \\
Gemini 4 Argon & 3/3 & 0.65--0.83 & 2--5 & 138--180 & 1.93--3.04 \\
Kimi K3 & 2/3 & 0.83--1.31 & 6--12 & 133--172 & 2.63--4.93 \\
Qwen3.8-max-0902 & 3/3 & 0.53--0.82 & 3--5 & 195--256 & 3.74--6.23 \\
DeepSeek-V4-Pro & 3/3 & 0.58--0.87 & 6--10 & 203--403 & 0.73--1.81 \\
DeepSeek-V4.1-Flash-1017 & 3/3 & 0.07--0.74 & 6--7 & 164--176 & 0.10--0.13 \\
\bottomrule
\end{tabular}
 \end{table}

\textbf{Completion.} Every session, completed or not, was scored at its end with $J$ of \cref{sec:protocol}. Twenty of the twenty-one sessions reached the three targets (\cref{tab:bench}); the one session that did not, Kimi K3 session 2, spent its 12 submissions at $J$ = 1.31. No session reached the 8 h of the budget. \Cref{tab:bench} also gives the number of tasks each session submitted, the length its context reached and its cost. \Cref{fig:bench-tools} shows the tool calls of each session. Claude Opus 5.5 made the fewest calls, 40 to 65 per session, and DeepSeek-V4-Pro the most, 152 to 248. GPT-6.1-sol ended with the shortest context, 45{,}000 to 65{,}000 tokens, and DeepSeek-V4-Pro the longest, 203{,}000 to 403{,}000. DeepSeek-V4.1-Flash-1017 cost the least, USD 0.10 to 0.13 per session, and Qwen3.8-max-0902 the most, USD 3.74 to 6.23. Each session submitted 2 to 12 tasks and called \executecode{} 4 to 63 times. Every session but Kimi K3 session 2 looked up the documentation. The MCP tools took 40\,\% to 81\,\% of the calls of a session.

\begin{figure}[pos=t]
  \centering
  \includegraphics{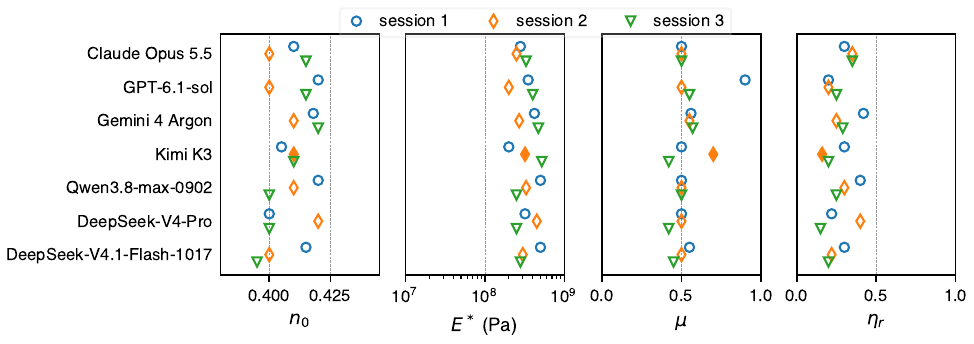}  \caption{The parameter set each session ended on, one strip per parameter; sessions 1 to 3 of each language model carry three markers, and the filled marker is the session that did not complete. The axis of each strip spans the range the message allows.}
  \label{fig:bench-solutions}
\end{figure}

\textbf{The calibrated parameters.} \Cref{fig:bench-solutions} shows the parameter set each session ended on. Nineteen of the twenty-one lie inside $n_0$ 0.395 to 0.420, $E^*$ $2.0$ to $5.2 \times 10^8$ Pa, $\mu$ 0.42 to 0.57 and $\eta_r$ 0.15 to 0.42, each span 15 to 38 percent of the range the message allows. The two outside it differ in $\mu$: GPT-6.1-sol session 1 completed with $\mu$ = 0.90, and Kimi K3 session 2, the session that did not complete, ended at $\mu$ = 0.70. Four parameters were fitted to three targets, each with a tolerance, so the targets are met in a region of the parameter space, not at one point.

The sessions reached the region by different routes. The output of a language model is random, so no two sessions ran alike, of one language model or of different language models. The sessions differed in the following.
\begin{itemize}
  \item The control of the walls. Ten sessions used the stress controller of the engine, and eleven wrote a servo of their own.
  \item The step between two iterations. Of the 62 steps in the twenty-one sessions, 19 changed one parameter and 43 changed two to four parameters at once.
  \item The trial before the first task. Fourteen sessions made 10 to 47 calls of \executecode{} before the first task, and seven made 1 to 7.
  \item The read of a running task. Fourteen sessions executed code snippets through \executecode{} while a task was running, and seven did not.
\end{itemize}

The strategies of the sessions were random, but the calibrated parameters of nineteen of the twenty-one sessions lie in one region, which spans 15\,\% to 38\,\% of each admissible range. The calibration is reproducible in this sense: the calibrated parameters agree between sessions, although the routes do not. The randomness of a language model acts on the route of a session; the three targets and the mechanics of the DEM model decide where the route can end. The record of the session of \cref{sec:campaign} shows how an agent stays on such a route: the agent chose each change of a parameter from the effect of the parameter on stiffness, strength and dilatancy.

\textbf{Mistakes and solutions.} The records of the sessions show the mistakes the agents made. \Cref{tab:mistakes} lists the mistakes by kind. Each row gives the sessions that made the mistake and states how the mistake was found and solved.

\begin{table}[pos=htbp]
\caption{The mistakes recorded in the twenty-one sessions, by kind, and their solutions. Sessions are numbered as in \cref{fig:bench-tools}.}
\label{tab:mistakes}
\small
\begin{tabular}{@{}>{\raggedright\arraybackslash}p{0.26\linewidth}>{\raggedright\arraybackslash}p{0.30\linewidth}>{\raggedright\arraybackslash}p{0.36\linewidth}@{}}
\toprule
Mistake & Sessions & Solution \\
\midrule
The agent wrote $q$ in Pa into the kPa column of the curve file. & DeepSeek-V4.1-Flash-1017 (1, 2)\newline DeepSeek-V4-Pro (2) & The agent found a peak deviator stress of 900{,}000 kPa. The agent converted the column. \\
\addlinespace
The agent reversed the sign of the volumetric strain, or of the wall stress. & DeepSeek-V4.1-Flash-1017 (1)\newline Claude Opus 5.5 (1) & The agent found a negative $D$, or negative wall stresses in the running preparation and interrupted the task. The agent flipped the sign. \\
\addlinespace
The agent passed an argument of the wrong type, or called a name that does not exist. & DeepSeek-V4.1-Flash-1017 (1, 2)\newline Kimi K3 (1, 2)\newline GPT-6.1-sol (2, 3)\newline Qwen3.8-max-0902 (2) & The agent read the engine's error from \texttt{check\_task\_status}. The agent found the right name in the documentation tools or by trial in \executecode{}. \\
\addlinespace
The loading walls did not move at the start of shear. & GPT-6.1-sol (1, 2, 3) & In sessions 2 and 3 the agent found zero wall velocity in a read of the running task through \executecode{} and interrupted the task; in session 1 the agent found a curve without axial strain. The agent corrected the controller. \\
\addlinespace
The lateral servo did not hold the confining stress at its target. & DeepSeek-V4.1-Flash-1017 (1, 2)\newline DeepSeek-V4-Pro (1) & The agent saw it in a read of the running shear through \executecode{}, or in the recorded lateral stress after the task. The agent corrected the servo. \\
\addlinespace
The agent never called the documentation tools. & Kimi K3 (2) & None. The session did not complete. \\
\bottomrule
\end{tabular}
\end{table}

Some of the mistakes were found in a running task, before the task ended and before $J$ was computed. In sessions 2 and 3 of GPT-6.1-sol, the agent read a wall velocity of zero through \executecode{} at the start of shear, interrupted the task and corrected the controller. In session 1 of Claude Opus 5.5, the agent read wall stresses of the wrong sign during the preparation of the specimen and interrupted the task. In session 3 of Kimi K3, the agent counted the contacts during shear, found a coordination number of 1.85 and half of the spheres with fewer than two contacts, and interrupted the task. Eight sessions of five language models interrupted a task in this way, on what the agent read through \executecode{}. These cases show the value of MCP tools designed for a live engine: with tools that only submit a simulation and wait for the result, the agent would have found each of these mistakes only after the task ended.

The one session that never called the documentation tools, Kimi K3 session 2, lost five of its twelve submissions to wrong calls before its first curve. This session is the one that did not complete. The two sessions of Kimi K3 that completed looked the documentation up 4 and 7 times. For this language model, reading the documentation reduced the wrong calls.

\subsection{A small language model in the engine's test pipeline}
\label{sec:pipeline}

This subsection reports a task other than the calibration, a short test of the execution tools given to a small language model in the test pipeline of YADE. A merge request to the YADE repository (\hyperref[sec:data]{Data availability}) adds a job in which Qwen3.5-4B, an open-weight language model of four billion parameters served on the CPU of the pipeline's runner, drives a YADE session through the bridge in the repository. The agent of the job is a short program that runs Qwen3.5-4B with the same five execution tools of \cref{sec:server}. The language model is asked to read a small model, to run a script as a task, to read the model while the task is running, to interrupt the task and to report the status of each task. The answer of the language model is checked against the state of the engine. The purpose of the job is to check that even a language model this small, on a CPU, can use the five execution tools of the MCP server. The job passes on the runner of the YADE pipeline.

 \section{Limitations}
\label{sec:limitations}

The interface has requirements on the engine, and the calibration task on which the interface was tested has limits.

\textbf{What the interface requires of the engine.} The bridge runs inside the engine's process, so the engine must embed a scripting language, Python in PFC and YADE, in which a server can listen for the agent's requests while the simulation runs. Two of the guarantees of \cref{sec:bridge,sec:semantics} rest on more than that: a read that completes within one step, and an interrupt that leaves the model intact at the step reached, both need a callback that the engine invokes in every cycle, the callback of PFC and the PyRunner of YADE. An engine that does not meet these requirements cannot serve the interface.

\textbf{What the calibration task did and did not test.} The tools were tested on one task, chosen to show whether an agent can drive the engine to a specified target, and the task has limits. The task was scored on three scalars, the secant stiffness, the peak strength and the dilatancy rate at the peak deviator stress, each with a tolerance. The rest of the test was not scored.
\begin{itemize}
  \item The axial strain at the peak deviator stress was reported but not scored.
  \item The shape of the curves before and after the peak was not a target.
  \item The post-peak softening and the critical state were not targets.
\end{itemize}
A parameter set that meets the three targets therefore reproduces three features of the test of \citet{fukushima1984}, not the whole test, and the sessions do not constitute a comparison of the DEM model with the full laboratory curves. The sessions were also not compared with a calibration by a modeler without an agent, in time or in the calibrated parameters. Such a comparison needs several modelers working on the same task, and is left to future work in human--computer interaction.
 \section{Conclusions}
\label{sec:conclusions}

This paper presented an MCP server for the agentic loop in discrete element simulation, in which a general-purpose AI agent and the modeler work on one engine process. Through the MCP server the agent looks up the engine's documentation, builds and runs a model, monitors the simulation, adjusts the model, and repeats until the task is done. The tools were derived from what the agent needs to do at the engine and were held to four principles: \emph{correctness}, \emph{coverage}, \emph{orthogonality} and \emph{simplicity}. A bridge inside the engine process runs the agent's code, and was built into a commercial DEM engine, PFC, and an open-source one, YADE. The tools were tested on one task, the calibration of a drained triaxial DEM model of dense Toyoura sand against laboratory data. The study showed the following.
\begin{enumerate}
  \item The documentation tools gave the agent the engine's manual without filling its context window. Two tools, a keyword query and a browse that discloses the manual level by level, kept the agent's code to what the engine documents (\emph{correctness}), reached every entry (\emph{coverage}), and needed no embedding model or vector database (\emph{simplicity}). The agent decided when to consult the documentation and which entries to read. The documentation guided the code the agent wrote, and an error from the engine drove the agent back to the documentation.
  \item The execution tools served both time scales of the work. The task tools served the long simulation, and the code snippet tool served the short call. The two did not overlap: a running task blocked neither a snippet nor a status check (\emph{orthogonality}). The agent submitted the long simulations as tasks in the background, checked on them, and stopped a task when needed. An interrupt left the model in the engine at the step where the task stopped. Code snippets ran both while the engine was idle and while a task was running, and the agent used them to read the evolving state of the simulation; all values in one read came from the same step (\emph{correctness}).
  \item The modeler and the agent worked on one engine process, the agentic loop of \cref{fig:architecture}. The modeler talked with the agent in the conversation, controlled the engine at its console, and watched the model in its graphical interface. The agent operated the engine through the MCP tools and read numbers and images from the engine. The modeler saw what the agent did, and the agent saw what the modeler did. The modeler watched the agent's simulations in the graphical interface as the simulations ran. What the modeler did at the console reached the agent with a tool result, and the agent acted on what the modeler had done. The study is thereby a step toward human--computer interaction in geotechnical simulation.
  \item A general-purpose agent completed the calibration through these tools, in PFC and in YADE. The agent read the documentation and tested commands on small models before submitting a task, diagnosed a failed simulation from the records of the simulation, and attributed each deviation from a target to a parameter, reasoning in the terms of soil mechanics. Seven language models then ran the same calibration on YADE without human intervention, and twenty of twenty-one sessions reached the tolerances. The sessions took different routes, and nineteen of the twenty-one parameter sets lie in one small region of the admissible ranges. In fourteen sessions the agent read a running task through a code snippet, and in eight of the fourteen the agent interrupted the task on what the read showed. These sessions show the value of MCP tools designed for a live engine. The one session that did not reach the tolerances was the only session that never called the documentation tools.
\end{enumerate}
 
\section*{Data availability}
\phantomsection\label{sec:data}
The two calibration sessions of \cref{sec:campaign,sec:yade} are archived at Zenodo \citep{demmcpdata2026}, \url{https://doi.org/10.5281/zenodo.23054301}. For each session the archive holds the opening message, the complete conversation between the modeler and the agent with every tool call and its result, the bridge log and the engine output of every task, the commands typed at the engine's console, the agent's notebook and scripts, and the curve of every iteration. The archive also holds the digitized test data, the measurements of the documentation tools in \cref{sec:docs}, and the records of the 21 sessions of \cref{sec:models}. 

The MCP servers and their bridges are open source under the MIT license: itasca-mcp \citep{itascamcp2026} and yade-mcp \citep{yademcp2026}, each installable from PyPI together with its bridge. The YADE bridge has also been submitted to the YADE repository as merge request 1187, \url{https://gitlab.com/yade-dev/trunk/-/merge_requests/1187}, where it is under review at the time of writing. A second merge request, 1216, \url{https://gitlab.com/yade-dev/trunk/-/merge_requests/1216}, adds the test job of \cref{sec:pipeline} to the YADE pipeline; it is a draft at the time of writing. The PFC session used itasca-mcp 0.10.1 with itasca-mcp-bridge 0.6.2, and the YADE session used yade-mcp 0.11.0 with yade-mcp-bridge 0.10.0. The agent in both sessions was Claude Code 2.1.283 \citep{claudecode2026} with the language model Claude Opus 5.5. The sessions of \cref{sec:models} used OpenCode 1.19.3 \citep{opencode2026} with yade-mcp 0.11.1, yade-mcp-bridge 0.10.0 and the YADE daily build of 2026-08-16. YADE is open source; PFC is commercial software of Itasca Consulting Group and requires a license to run.

\section*{Declaration of competing interest}
The authors declare that they have no known competing financial interests or
personal relationships that could have appeared to influence the work reported
in this paper.

\section*{Declaration of generative AI and AI-assisted technologies in the manuscript preparation process}
The simulations of the case study (\cref{sec:casestudy}) were run by general-purpose AI agents, which operated PFC and YADE autonomously through the MCP tools of \cref{sec:server}; the tools serve any agent that supports the protocol. This use of AI is the subject of the study, not an aid to it. A human modeler, one of the authors, set each task, watched each session at the engine and could stop it at any time. The authors reviewed and confirmed every result the agents reported: each score was recomputed with the scoring script, which the agents did not modify, and the accepted parameter set of each engine was rerun after its session and reproduced the curve the agent had recorded, row for row.

During the preparation of this work the authors used a general-purpose AI agent in order to draft and edit the text under the authors' direction, write the plotting scripts, check the references against their sources and assemble the data archive. The authors decided the research questions, the design of the tools and of the case study, and every claim and conclusion of the paper. After using this tool, the authors reviewed and edited the content as needed and take full responsibility for the content of the published article.

\section*{Acknowledgements}
The first author acknowledges the financial support provided by the Japan Science and Technology Agency (JST) SPRING program (Grant Number JPMJSP2110). The authors thank Bruno Chareyre for testing the MCP server and the bridge in YADE with an AI agent, and Janek Kozicki for suggestions on the test pipeline and on the documentation of the bridge. The authors also thank the developers of YADE for the discussion of the merge requests and for access to the test pipeline of YADE.

\printcredits

\bibliographystyle{cas-model2-names}
\bibliography{refs}

\end{document}